\documentclass[10pt]{article}

\usepackage{fullpage}
\usepackage{setspace}
\usepackage{parskip}
\usepackage{titlesec}
\usepackage[section]{placeins}
\usepackage{xcolor}
\usepackage{breakcites}
\usepackage{lineno}
\usepackage{hyphenat}

\PassOptionsToPackage{hyphens}{url}
\usepackage[colorlinks = true,
            linkcolor = blue,
            urlcolor  = blue,
            citecolor = blue,
            anchorcolor = blue]{hyperref}
\usepackage{etoolbox}

\renewenvironment{abstract}
  {{\bfseries\noindent{\abstractname}\par\nobreak}\footnotesize}
  {\bigskip}

\titlespacing{\section}{0pt}{*3}{*1}
\titlespacing{\subsection}{0pt}{*2}{*0.5}
\titlespacing{\subsubsection}{0pt}{*1.5}{0pt}

\usepackage{authblk}

\usepackage{graphicx}
\usepackage[space]{grffile}
\usepackage{latexsym}
\usepackage{textcomp}
\usepackage{longtable}
\usepackage{tabulary}
\usepackage{booktabs,array,multirow}
\usepackage{amsfonts,amsmath,amssymb}
\newif\iflatexml\latexmlfalse

\AtBeginDocument{\DeclareGraphicsExtensions{.pdf,.PDF,.eps,.EPS,.png,.PNG,.tif,.TIF,.jpg,.JPG,.jpeg,.JPEG}}

\usepackage[utf8]{inputenc}
\usepackage[english]{babel}
\usepackage{float}
\usepackage[margin=1.5in]{geometry}
\usepackage{csquotes}
\usepackage[style=apa,backend=biber, eprint=true]{biblatex} 
\usepackage{subfig}

\title{A Review of the State-of-the-Art on Tours for Dynamic Visualization of High-dimensional Data\\
}

\author{
 
  Stuart Lee\\
  Department of Econometrics and Business Statistics, Monash University \\
  \href{mailto:stuart.a.lee@monash.edu}{\nolinkurl{stuart.a.lee@monash.edu}}\\
  
  ~\\ \and 
 
  Dianne Cook\\
  Department of Econometrics and Business Statistics, Monash University \\
  \href{mailto:dicook@monash.edu}{\nolinkurl{dicook@monash.edu}}\\
  
  ~\\ \and 
 
  Natalia da Silva\\
  Instituto de Estadística (IESTA), Universidad de la República \\
  \href{mailto:natalia@iesta.edu.uy}{\nolinkurl{natalia@iesta.edu.uy}}\\
  
  ~\\ \and 
 
  Ursula Laa\\
  Institute of Statistics, University of Natural Resources and Life Sciences \\
  \href{mailto:ursula.laa@boku.ac.at}{\nolinkurl{ursula.laa@boku.ac.at}}\\
  
  ~\\ \and 
 
  Nicholas Spyrison\\
  Faculty of Information and Technology, Monash University \\
  \href{mailto:nicholas.spyrison@monash.edu}{\nolinkurl{nicholas.spyrison@monash.edu}}\\
  
  ~\\ \and 
 
  Earo Wang\\
  Department of Statistics, The University of Auckland \\
  \href{mailto:earo.wang@auckland.ac.nz}{\nolinkurl{earo.wang@auckland.ac.nz}}\\
  
  ~\\ \and 
 
  H. Sherry Zhang\\
  Department of Econometrics and Business Statistics, Monash University \\
  \href{mailto:huize.zhang@monash.edu}{\nolinkurl{huize.zhang@monash.edu}}\\
  
  ~\\ \and 

}

\date{}

\begin{document}

\maketitle

\vspace{-1em}

\begin{abstract}
This article discusses a high-dimensional visualization technique called the tour, which can be used to view data in more than three dimensions. We review the theory and history behind the technique, as well as modern software developments and applications of the tour that are being found across the sciences and machine learning.

\end{abstract}

\hypertarget{introduction}{%
\section{Introduction}\label{introduction}}

Data commonly arrives with more than two measured variables, which makes it more complicated to plot on a page. With multiple variables, especially if there is some association between variables, this would be called multivariate or high-dimensional data. When the variables are all numeric, or quantitative, visualization often relies on some form of dimension reduction. This can be done by taking linear projections, for example, principal component analysis \autocite{Hotelling1933-of} or linear discriminant analysis \autocite{Fisher1936}. It is also common to reduce dimension with nonlinear techniques like multidimensional scaling (MDS) \autocite{Kruskal1964-do} or t-Distributed Stochastic Neighbor Embedding (t-SNE) \autocite{Van_der_Maaten2008-qa}.

\begin{figure*}[!h]
\centerline{\includegraphics[width=0.95\textwidth]{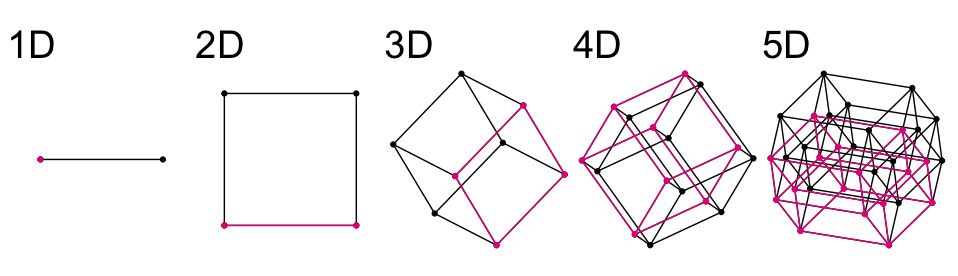}}
\caption{Illustrating what is meant by "high-dimensional" in this paper, and a "linear projection". From a sequence of increasing dimension cubes, from 1D to 5D, as wireframes, it can be seen that the as dimension increase by 1, the cube doubles. }
\label{fig:cubes}
\end{figure*}

The term ``high-dimensional'' here means Euclidean space. Figure \ref{fig:cubes} shows a way to imagine this. It shows a sequence of cube wireframes, ranging from 1D through to 5D, where beyond 2D is a linear projection of the cube. As dimension increase, a new orthogonal axis is added. For cubes, this is achieved by doubling the cube: a 2D is two 1D cubes, a 3D is two 2D cubes, and so forth.

The focus of this review will be on visualizing high-dimensional numerical data using linear projections, in particular, as provided by the grand tour \autocite{Asimov1985-xr,Buja2005-cx}. The reason being that it is not feasible to adequately review the very large area of visualizing high-dimensions, and there have been numerous developments in tours recently. An overview of the technique, and new modifications is provided, along with how these techniques can be used in a variety of applications.

A tour can be considered to be a dynamic graphic, because it shows a smooth sequence of projections over time, ideally with controls that allow stopping, reversing, changing direction, or going forward again. It can be useful to embed a tour into an interactive graphics system, where plots can be queried and elements highlighted (see for e.g. \textcite{Swayne2003} or \textcite{Tierney1991}). To create the smooth sequence, a geodesic interpolation is computed between consecutive frames. It allows the viewer to extrapolate from the low-dimensional to shapes corresponding to multivariate distribution, and is particularly useful for detecting clusters, outliers and non-linear dependence.

While tours are invaluable for assessing the geometry of data, they are by no means the only technique available for visualizing structure in high dimensional data. An early technique proposed for assessing pairwise relationships between variables is the scatterplot matrix (SPLOM) \autocite{Tukey1981-ov,Chambers1983-dh,Tukey1983-fj,Carr1984-xw,Becker1987-dw}. The SPLOM allows the viewer to assess correlation structure but does not scale to large numbers of variables. Similarly, parallel coordinates plots (PCP) can be used to explore correlation and collinearity \autocite{Inselberg1985-kf,Wegman1990-dy}. By placing multiple variables side by side in a PCP higher order structure like clustering or lower dimensional embeddings are revealed, however the ordering of variables along the axis changes what can be learned. Another display that relies on variable ordering is the heatmap, which is widely used to visualize cluster structure in bioinformatics. \textcite{Wilkinson2009-pj} provides a comprehensive history of this display in the social and natural sciences.

\begin{figure*}[!ht]
   \centering
         \includegraphics[width=0.45\textwidth]{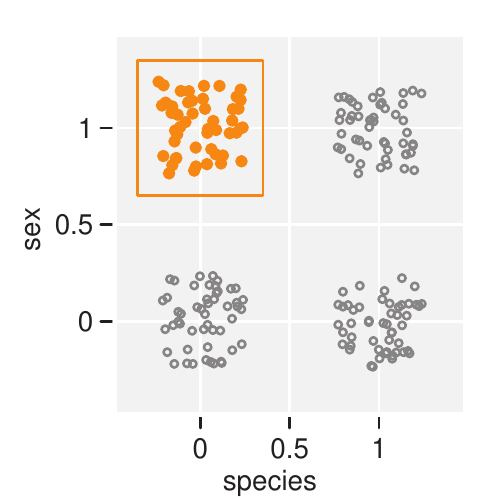}
         \includegraphics[width=0.45\textwidth]{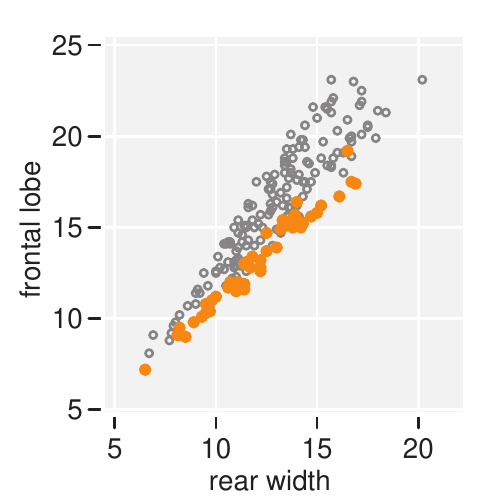}
\caption{When a user brushes data on the left panel, points that fall inside the region are highlighted in orange. The view on the right responds by highlighting the corresponding points. Adapted from \cite{Cook2007-be} Figure 2.12.}
\label{fig:brushing}
\end{figure*}

All of the aforementioned techniques can be enhanced through the use of interactivity.
By combining views with interactive elements like tool tips or highlighting the analysts is able to quickly interrogate interesting features of the data. One particularly important interaction technique in the history of statistical graphics is called brushing \autocite{Becker1987-dw}. When brushing an analyst drags their mouse over the view which results in a region being drawn onto the canvas. When there are multiple views present the act of brushing can be thought of as a database query; points that fall inside the brush can be used to highlight or filter data on adjacent views (Figure \ref{fig:brushing}). This technique is particularly useful when combined with the tour (Section \ref{sec:spin}).

\begin{figure*}[!t]
         \includegraphics[width=0.371\textwidth]{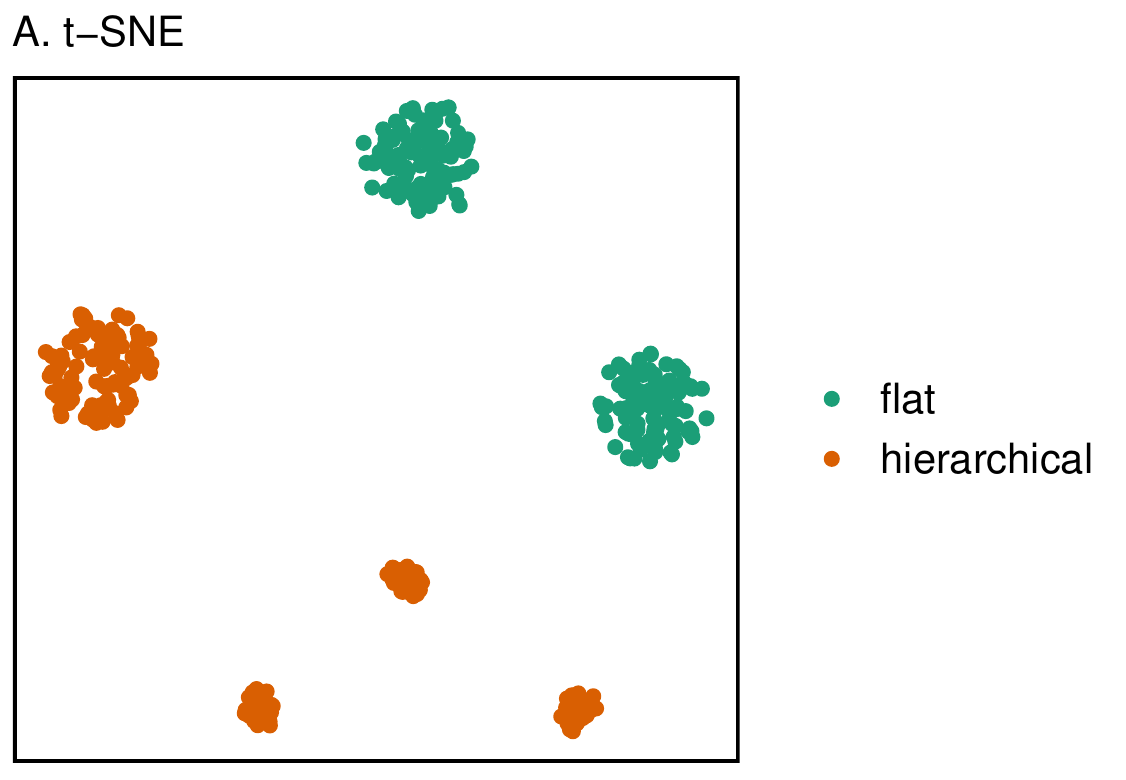}
     \vfill
   \begin{minipage}{\linewidth}     
   \centering
         \includegraphics[width=0.24\textwidth]{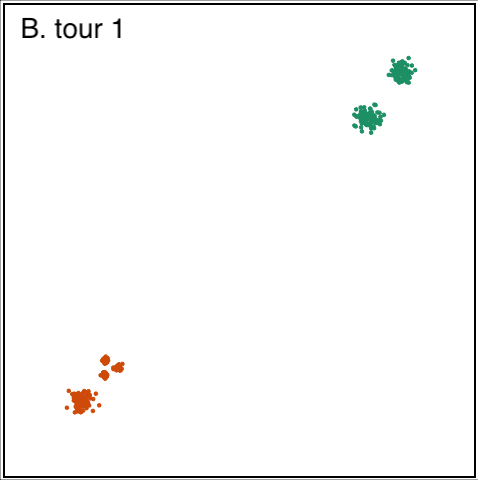}
         \includegraphics[width=0.24\textwidth]{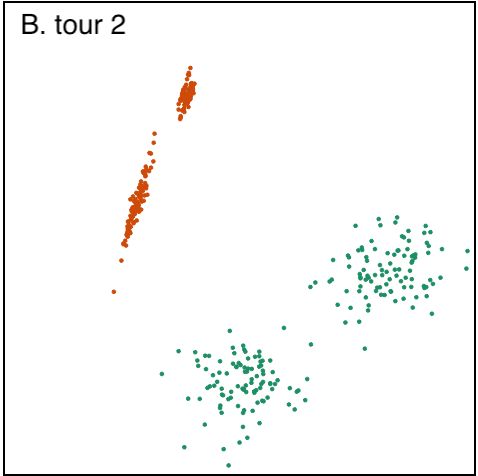}
         \includegraphics[width=0.24\textwidth]{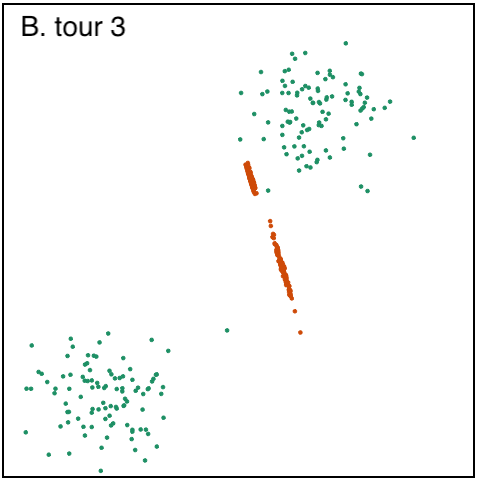}
         \includegraphics[width=0.24\textwidth]{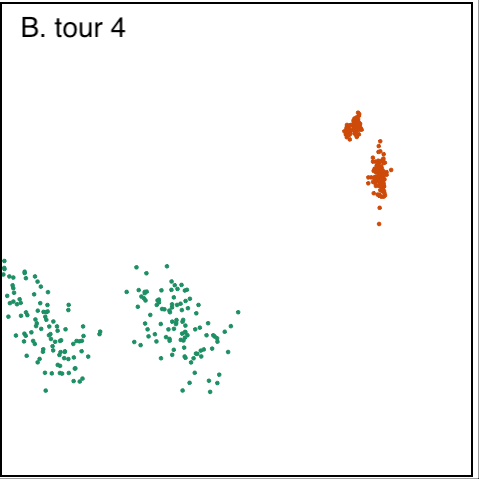}
   \end{minipage}%
\caption{Comparison of structure perception between nonlinear dimension reduction from 10-d, using t-SNE (A) and a tour (B). There are six clusters, as seen in the t-SNE view, but the relative distance between the clusters is extremely varied. This can be seen in sample of tour projections shown. The two green clusters are (almost) spherical in shape, and very distance from the orange clusters. Three of the orange clusters are very close to each other (just visible in B.1), and all orange clusters are elliptical. The tour provides a more accurate rendering of the clusters in the high-dimensional space, and complements what is learned from the dimension reduction.}
\label{fig:tsne-tour}
\end{figure*}

Nonlinear dimension reduction techniques such as t-SNE and Uniform Manifold Alignment and Projection (UMAP) have become very popular in recent years, primarily for the ability to capture cluster structure in a succinct visual summary \autocite{Van_der_Maaten2008-qa,mcinnes2020umap}. However, it is only a summary, and it likely involves substantial warping of the original data space. Using the tour along with these techniques can illuminate the nature of the warped space, and reveal other structure lost in the dimension reduction. Figure \ref{fig:tsne-tour} illustrates the difference in what can be learned from a tour in comparison with nonlinear dimension reduction using t-SNE. The 10-dimensional data comes from \textcite{Rauber2009-kk}. The t-SNE view (plot A) shows six clearly separated clusters, all spherical with different sizes. The tour shows that the clusters do not look like this in the full data space. The clusters are at various distances apart and very different sizes, as can be seen from the four projections from a tour. The two green clusters are large and almost spherical, and far from the orange clusters. The orange clusters have one larger one, and three smaller, very close to each other. All of these are elliptical, which means that they have very little variability, actually no variability if you watch the full tour, in some of the 10 dimensions. This gives some deeper perspective to what is learned from t-SNE, and illustrates what the t-SNE dimension reduction has done: it has found small gaps between points and expanded these gaps to yield the representation. It should be noted, though, that the methods (t-SNE and tour) complement each other. The t-SNE view gives a clear indication of six clusters, which may have been overlooked on initial viewing with a tour. With this information is the invitation to look closer at the data in the tour, to see, that yes, indeed, there are three, tiny, tiny clusters very close to each other.

The rest of the review is structured as follows: Section \ref{notation} defines the notation and components of a tour displays. By its nature a tour is most effective to analyst when combined with interactivity; Section \ref{interaction} reviews the components of user interfaces for manipulating tour views, including manual tours (Section \ref{tours}) which are useful to test structure of selected features. The implementation of tour paths in statistical software is reviewed in Section \ref{software}. Section \ref{applications} shows the diverse applications of the tour in the natural sciences, machine learning and applied statistics. Finally, Section \ref{discussion} discusses future research directions for tours.

\hypertarget{notation}{%
\section{Tours for high dimensional visualization}\label{notation}}

\hypertarget{notation-1}{%
\subsection{Notation}\label{notation-1}}

\begin{figure*}
   \centering
         \includegraphics[width=\textwidth]{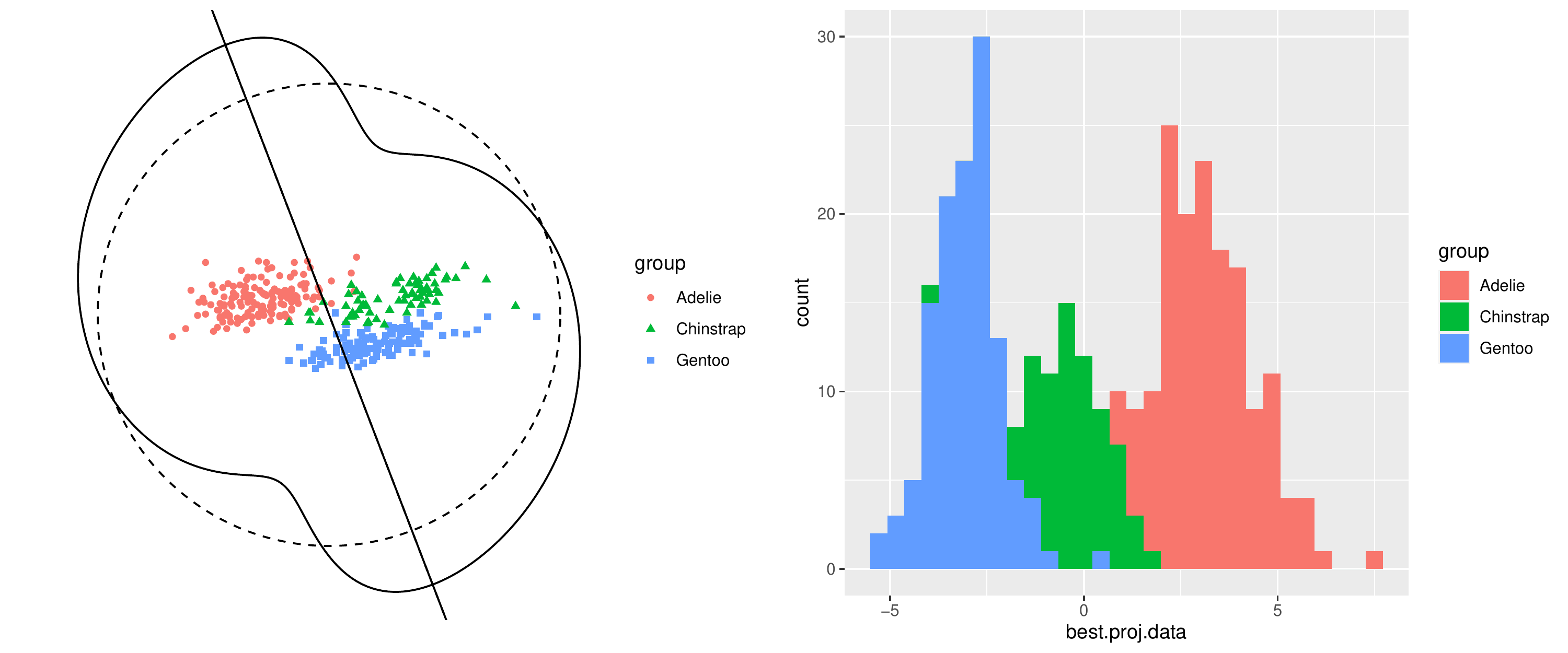}
         \includegraphics[width=\textwidth]{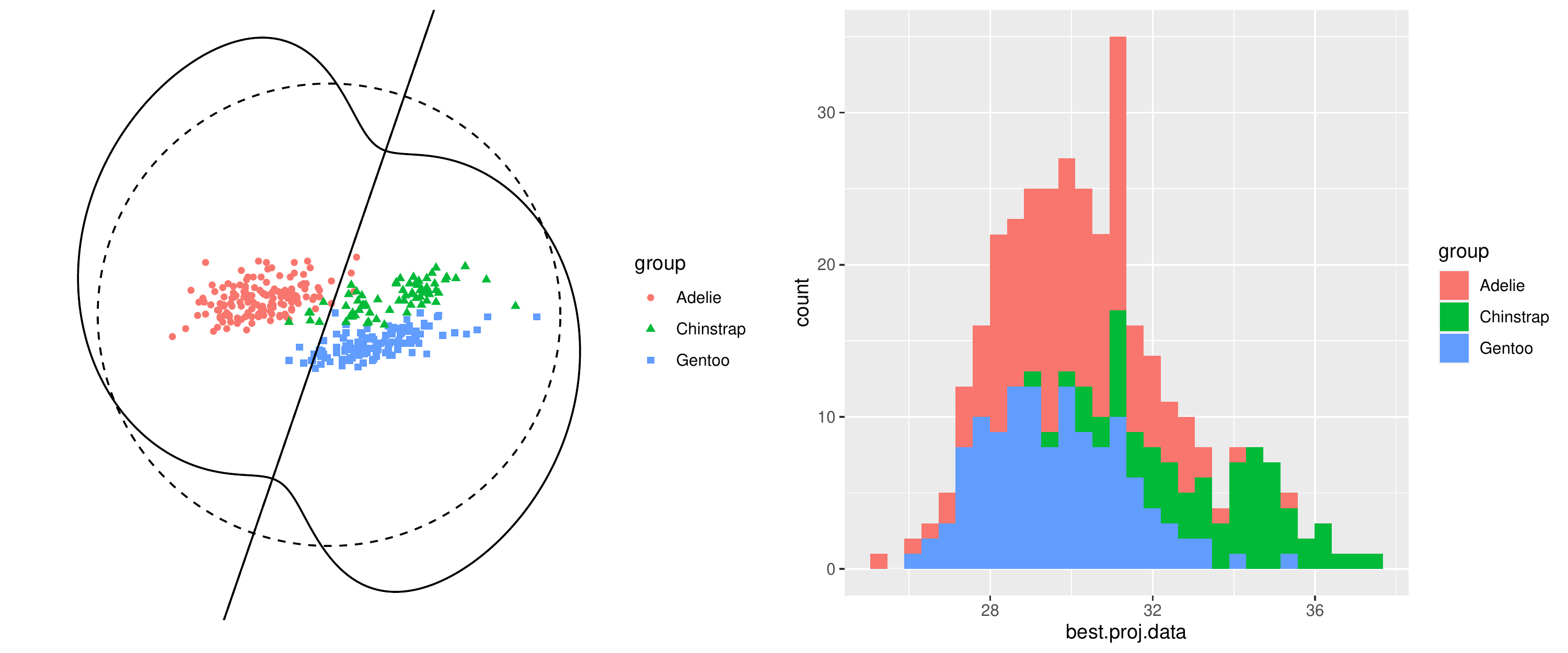}
\caption{Huber's plot (left plots) of the penguins data (bill length and depth only). The solid line represents how well the projection onto different directions, separates the three species, and the dashed circle is a reference guide. The histogram (right plots) is the 1D projection of the data onto the direction outlined as the solid line in the Huber's plot. Adapted from \cite{Lee2005-gl}, Figure 1.}
\label{fig:notation-huber}
\end{figure*}

When using a tour, a sequence of \(d\)-dimensional linear projections are obtained from a \(p\)-dimensional space, where \(d \ll p\). Let \(X_{n \times p}\) be the data matrix, consisting of \(n\) observations and \(p\) variables, whose projections are of interest. A projection basis, \(A_{p \times d}\), is matrix that characterizes the direction from which the data are projected and needs to satisfy an orthonormality condition, which requires that each column in \(A\) has unit length and are perpendicular. With a data matrix and a projection basis, a projection of the data can be defined as \(Y = X \cdot A\). Figure \ref{fig:notation-huber} shows two examples of low-dimensional projections of the palmer penguins data \autocite{penguins} in a Huber plot \autocite{huber1990} and a histogram.

\begin{figure*}
   \centering
         \includegraphics[width=\textwidth]{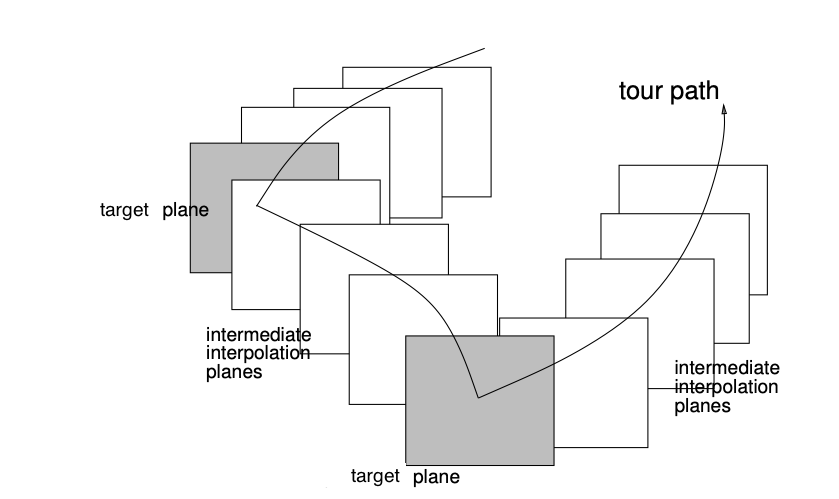}
\caption{Illustration of a tour path, indicating target planes, and interpolation planes. Adapted from \cite{Buja2005-cx}, Figure 1.}
\label{fig:tour-path}
\end{figure*}

\hypertarget{finding-targets}{%
\subsection{Finding targets}\label{finding-targets}}

A tour path, that is a sequence of projections, can be generated by geodesic interpolation between a set of target planes (Figure \ref{fig:tour-path}). Different methods for choosing target planes provide different tour paths. The grand tour is generated using randomly selected target. A guided tour is generated by choosing particularly structured projection from a projection pursuit optimization. The little tour uses all variable bases as targets, and a local tour rocks back and forth from a particular plane to randomly chosen targets in a small neighborhood.

\begin{figure}
\centering
\includegraphics[width=\textwidth]{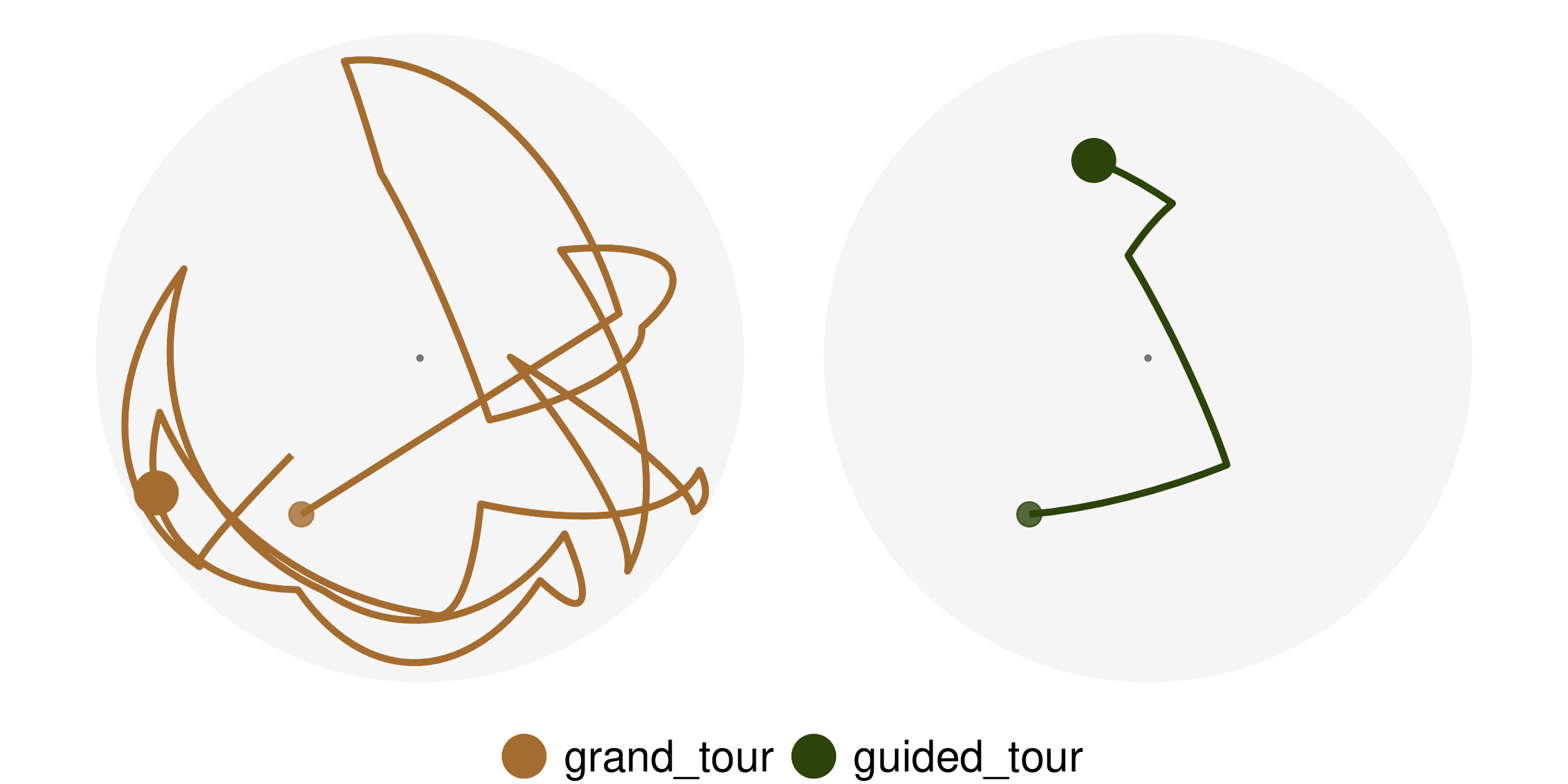}
\caption{A comparison of guided and grand tour path in the PCA reduced 2D space. The start point of tour paths is smaller and lighter than the end point. Guided tour optimises the index value iteratively and finishes the search quickly while grand tour wanders in the parameter space to look at possible interesting projections.}
\label{fig:notation-target-gen}
\end{figure}

Figure \ref{fig:notation-target-gen} shows a representation of 1D tour paths of 5D data, drawn on a PCA space. A grand tour path is on the left and a guided tour path is on the right. The grand tour path should be more wide-spread because it is attempting to show all possible projections, if left to run long enough. The guided tour path should be short, as the projection pursuit optimization zeroes on the optimally interesting projection.

There are other tour types which don't fit this style of tour. The manual tour \autocite{cook_manual_1997} allows the user to change the projection coefficients manually, to rotate a variable into and out of a project, and is discussed later. The recently developed slice tour \autocite{Laa2020-js} can be applied to any of the above tour types. It displays a slice through the orthogonal space, as opposed to a projection, and is explained in more details later.

\hypertarget{geodesic-interpolation}{%
\subsection{Geodesic interpolation}\label{geodesic-interpolation}}

The smooth progression in the tour path is due to geodesic interpolation between the target planes. This takes into consideration two important aspects: (1) maintains the orthonormality of the projection bases, (2) contains all the rotation to be between planes, not a particular basis in any plane. The first is clearly important because it ensures that we a looking at low-dimensional projection of the data always. The latter is harder to explain, but really important from a visual perspective. It stops any within-plane spin, and could be considered stabilizing the view. More details explanation of this can be found in \textcite{Buja2005-cx}.

\begin{figure*}[ht]
\centerline{\includegraphics[width=0.3\textwidth]{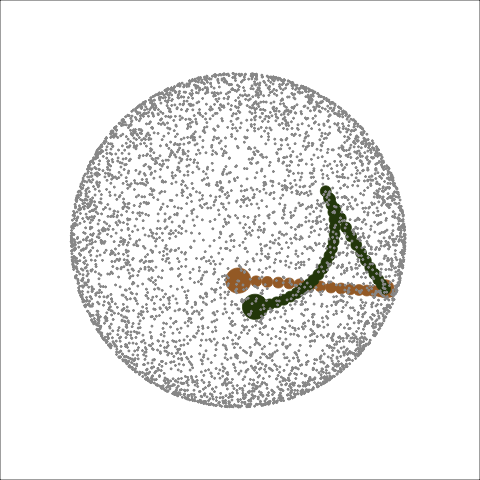}
\includegraphics[width=0.3\textwidth]{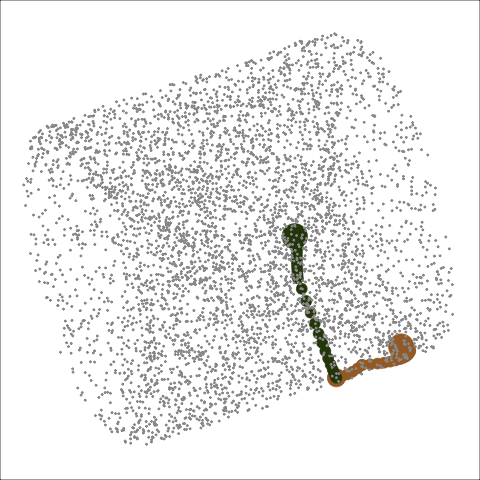}
\includegraphics[width=0.3\textwidth]{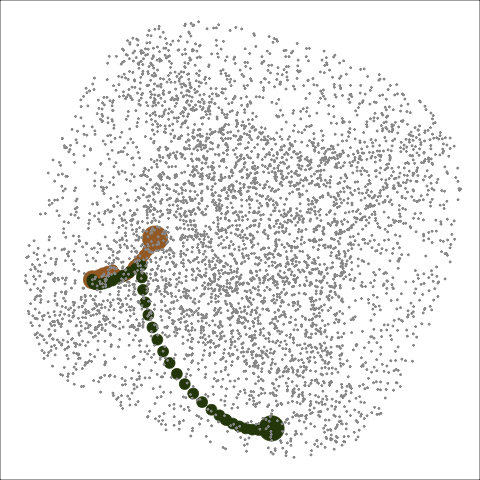}}
\centerline{
\includegraphics[width=0.3\textwidth]{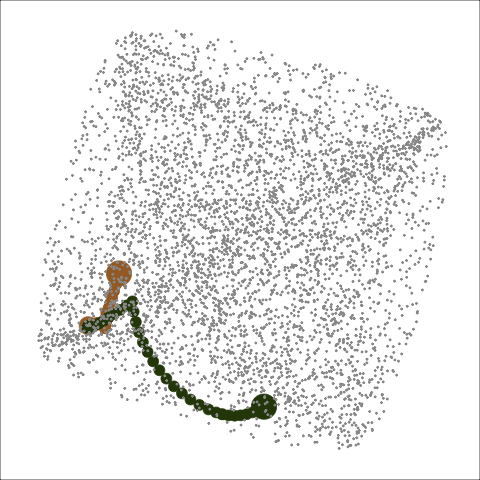}
\includegraphics[width=0.3\textwidth]{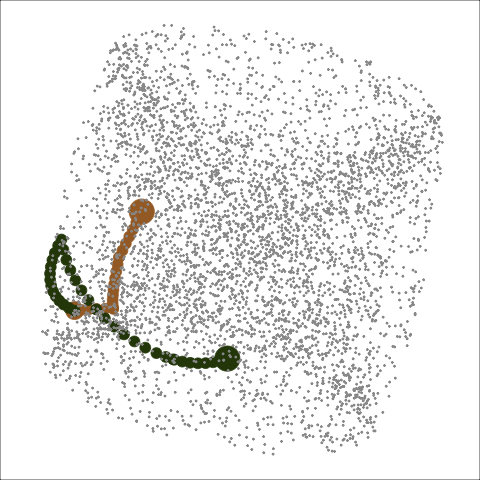}
\includegraphics[width=0.3\textwidth]{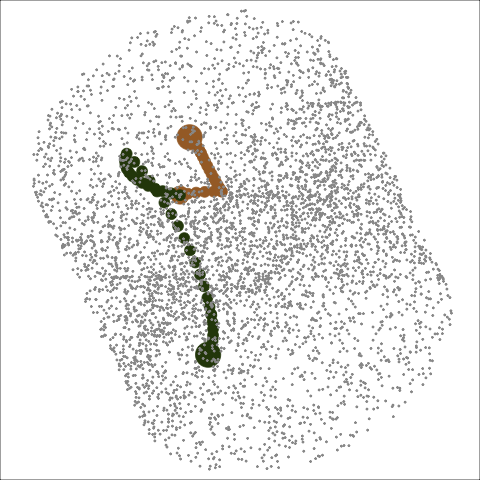}
}
\caption{Still views of a tour showing the space of possible two-dimensional projections as a torus shape in grey, and the trace of two short paths obtained via geodesic interpolation traced in color.}
\label{fig:interpolation-2d}
\end{figure*}

Figure \ref{fig:interpolation-2d} shows tour paths of 2D projections of 6D data. The space of all tour paths is a high-dimensional torus, as represented by the gray points. The tour paths are shown in green and orange, and each dot indicates a projection in the sequence. The plots in this figure could be considered to be a tour looking at itself, because each plot is a selected view from a tour of the torus with the paths overlaid.

\hypertarget{display}{%
\subsection{Display}\label{display}}

\begin{figure}
\centering
\includegraphics[width=0.3\textwidth]{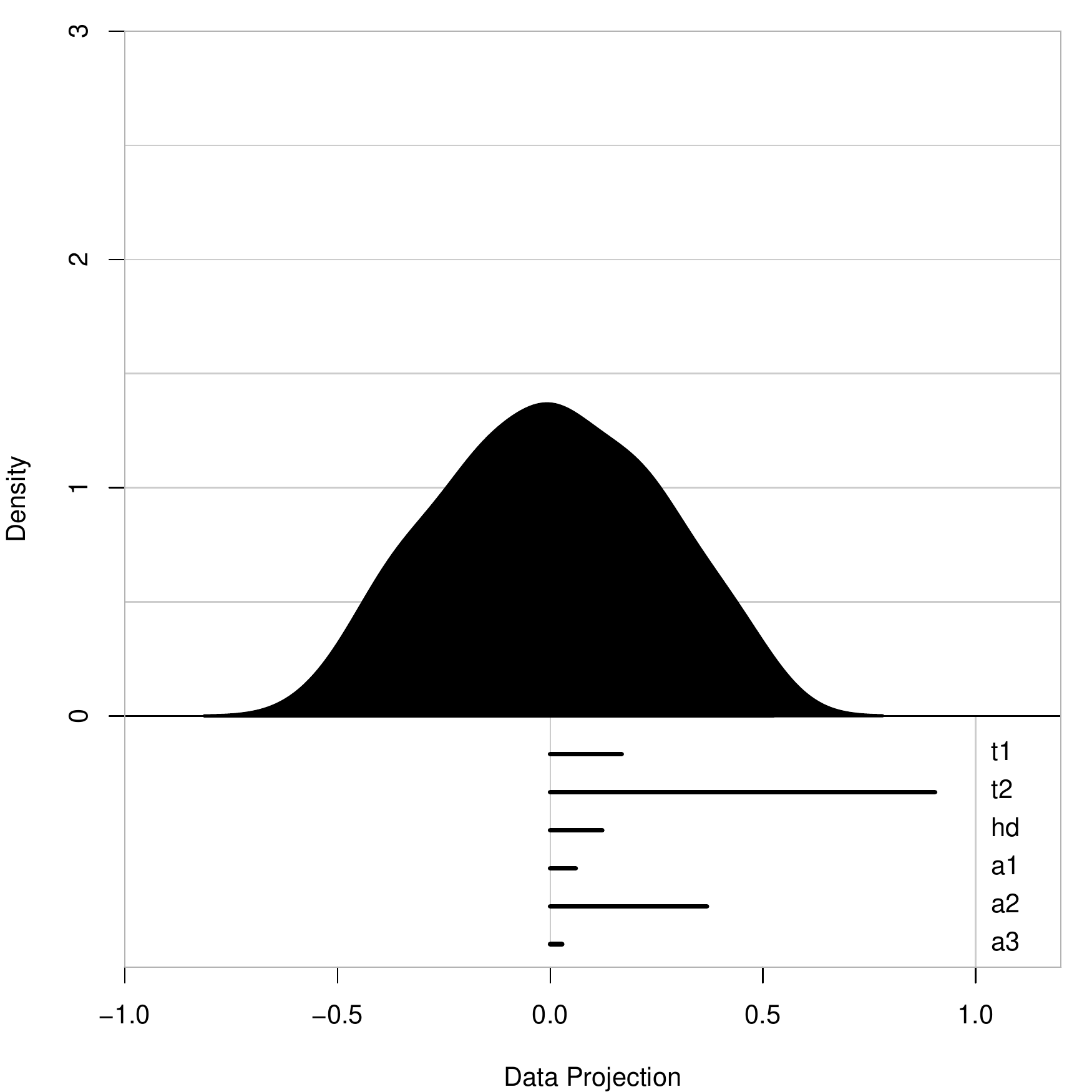}
\includegraphics[width=0.3\textwidth]{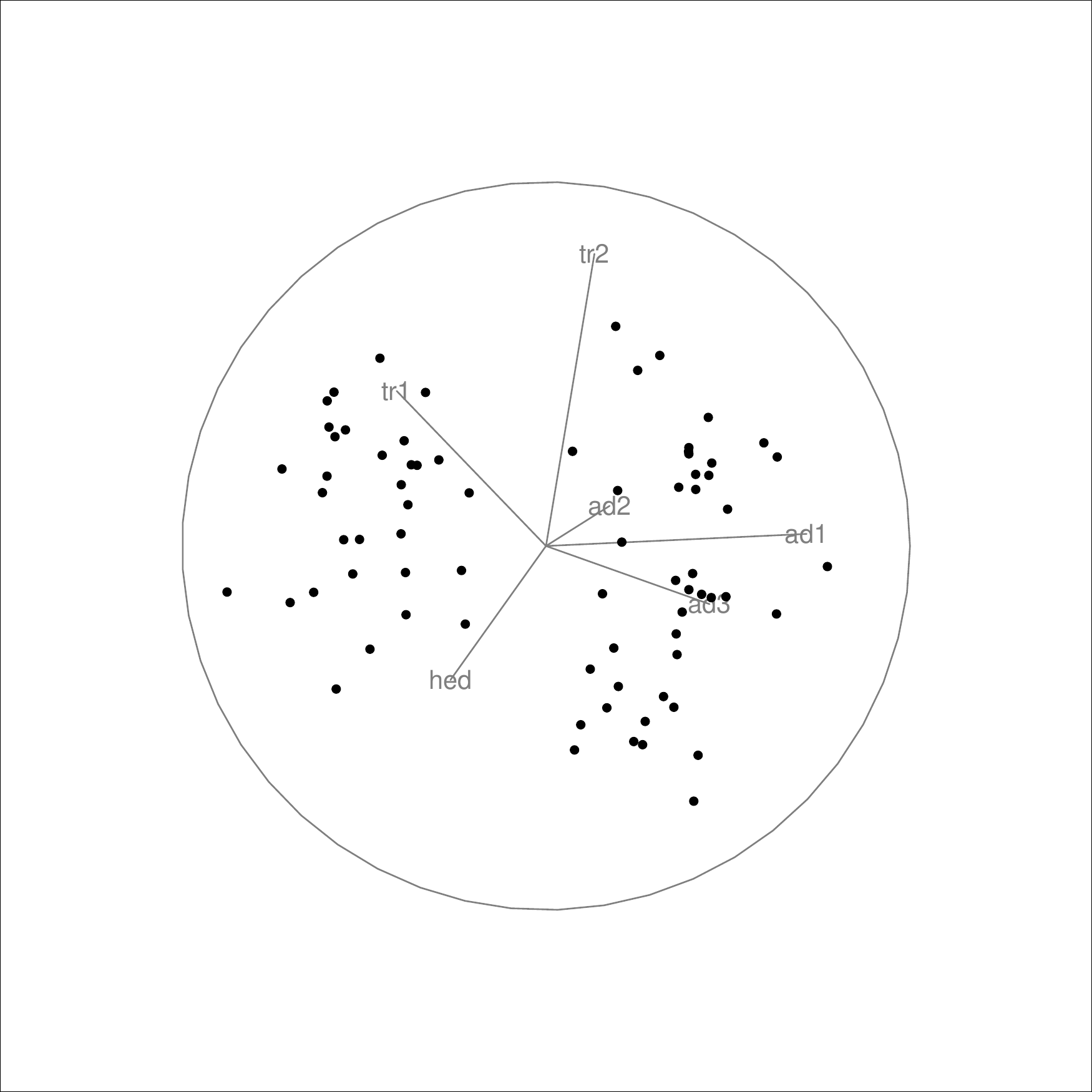}
\includegraphics[width=0.3\textwidth]{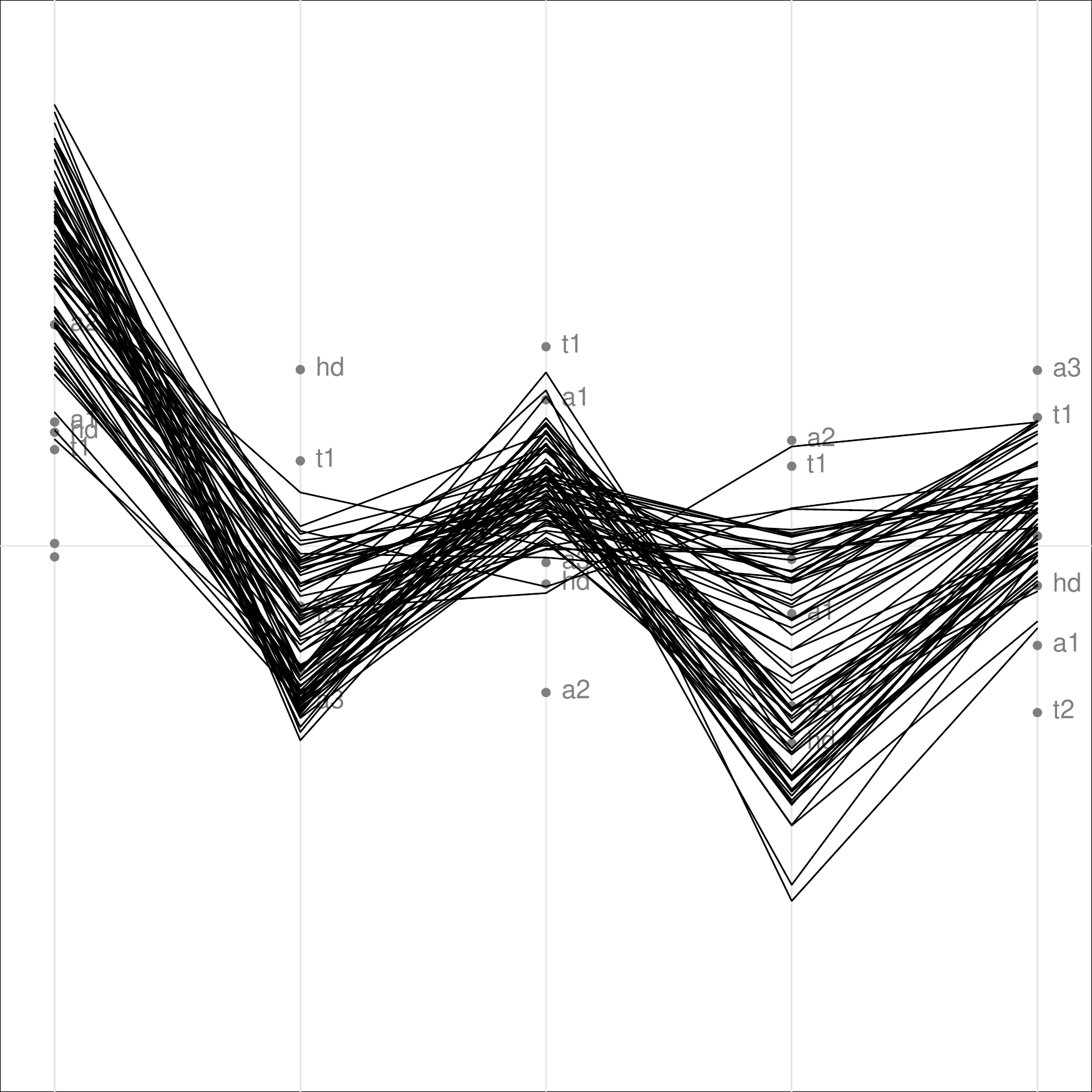}
\caption{Various displays of the projections: 1D display in histogram (left), 2D display in scatterplot (middle), and 5D display in parallel coordinate plot (right).}
\label{fig:display}
\end{figure}

Various displays are available to show projections in 1D, 2D and higher dimensions. A 1D projection displays the data in a histogram analogous to a shadow puppet being projected onto a wall, while 2D projections are displayed as scatterplots. Higher dimensional projections can be shown with multivariate displays like SPLOMs or parallel coordinates. Figure \ref{fig:display} shows a variety of displays for projections and these displays are useful to show the non-normal distributions in the projected data, clustering structure, and multivariate relationship between variables.

\hypertarget{slices-sections-and-projections}{%
\subsection{Slices, sections and projections}\label{slices-sections-and-projections}}

Interactive systems like GGobi \autocite{Swayne2003} provide the option of adding sectioning information via linked brushing. Sectioning means we select points that fall in a section of the full parameter space, for example for the purpose of highlighting them in a tour display. This often reveals complementary information, and combining sections and projections can for example provide insights into the dimensionality of a data structure \autocite{Furnas1994-dz}.

Sectioning a high-dimensional space in systematic manner (without interactive selection of data sections) is challenging, since there is a lot of freedom in choosing a section. One approach is to define sections based on projection planes, we refer to such sections as ``slices'' of the data. This is implemented in the slice tour display \autocite{Laa2020-js}, and uses the orthogonal distance of each data point from the current projection plane (typically placed such that it passes through the data mean) to highlight points that are nearby and fade out points further from the plane.

\begin{figure*}[ht]
\centerline{\includegraphics[width=0.3\textwidth]{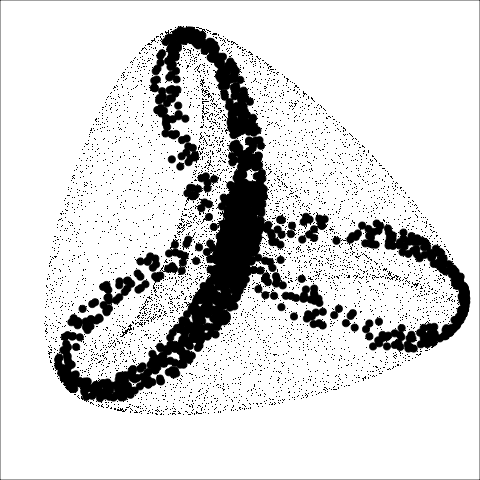}
\includegraphics[width=0.3\textwidth]{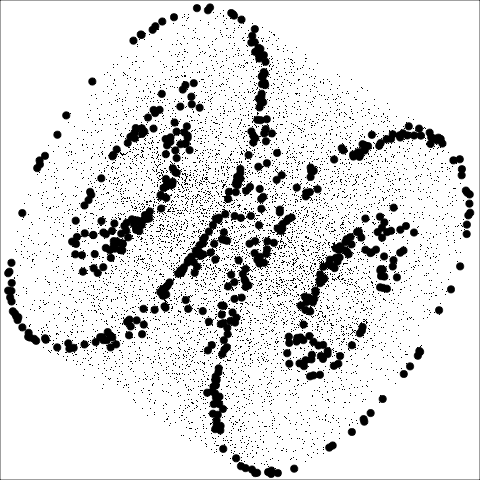}
\includegraphics[width=0.3\textwidth]{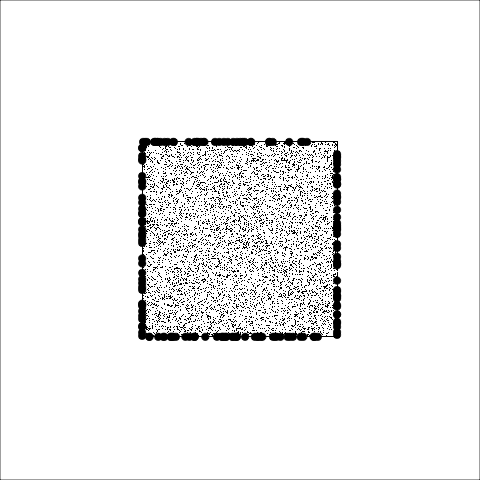}}
\centerline{
\includegraphics[width=0.3\textwidth]{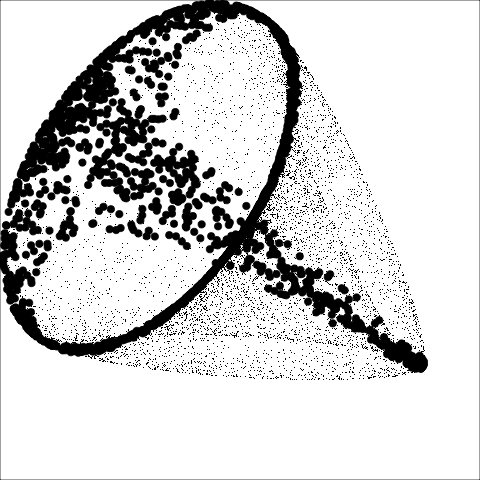}
\includegraphics[width=0.3\textwidth]{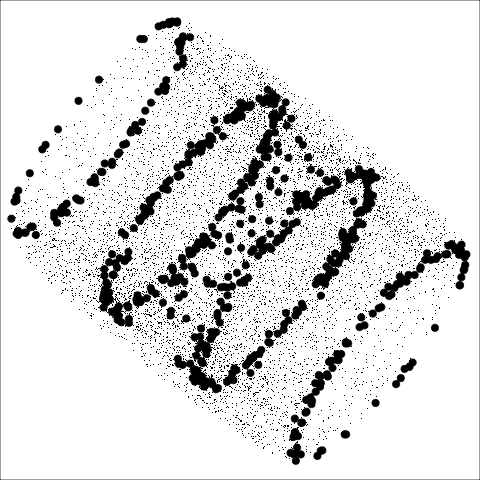}
\includegraphics[width=0.3\textwidth]{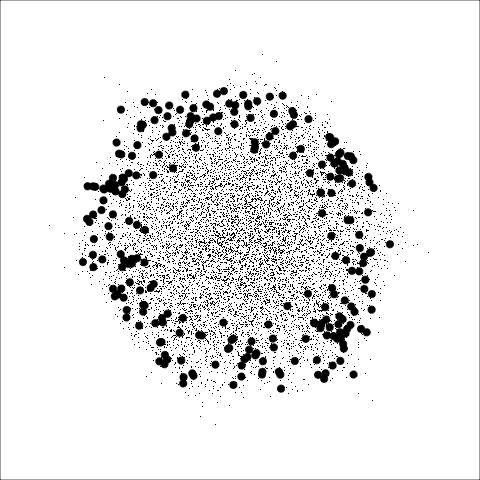}
}
\caption{Still views of the slice tour display showing geometric shapes: Roman surface (left), 4D torus (middle) and a 6d cube (right). Adapted from \cite{Laa2020-ni}, Figure 4.}
\label{fig:tour-slice}
\end{figure*}

The slice tour can reveal concave or non-linear structures obscured in projections, as well as small structures hidden near the center of a distribution. Figure \ref{fig:tour-slice} shows snapshots of the slice tour of points distributed on the surface of geometric shapes. Along with projections, an index function can be used to select target planes to show interesting slices of the data, defined as a section pursuit guided tour \autocite{Laa2020-ni}.

\hypertarget{transformations-on-projections}{%
\subsection{Transformations on projections}\label{transformations-on-projections}}

\begin{figure}
\centering
\includegraphics[width=0.3\textwidth]{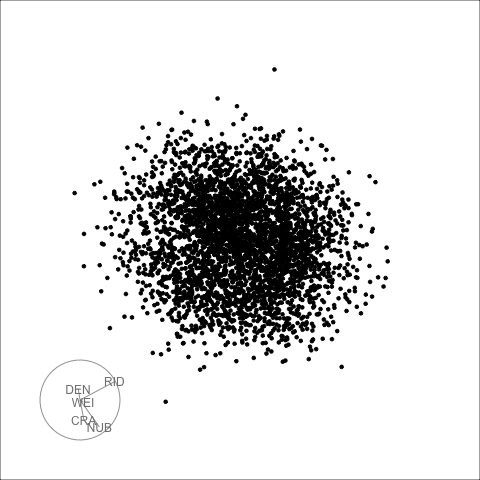}
\includegraphics[width=0.3\textwidth]{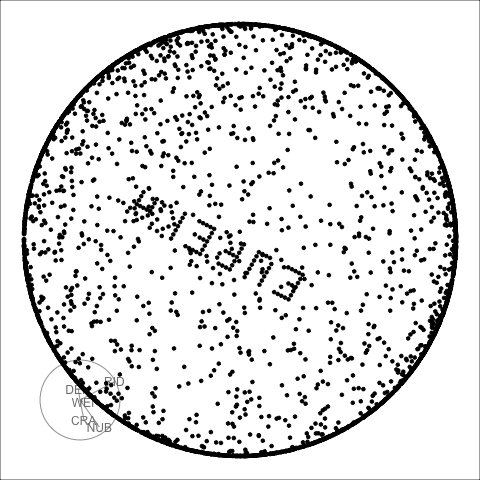}
\includegraphics[width=0.3\textwidth]{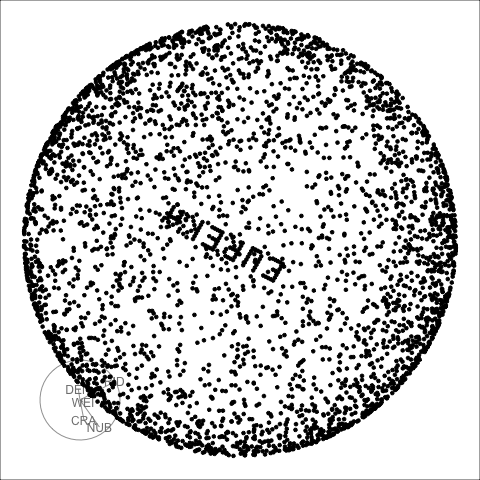}
\caption{Snapshots of the same projection from the sage tour on the pollen data, illustrating the different tuning options. Left: using the default settings for the transformation results in a similar view as found in projections without the sage transformation. Middle: using the tuning parameter $R$ for better resolution near the center reveals the word "EUREKA". Right: using the $\gamma$ tuning parameter can also reveal the structure. Adapted from \cite{Laa2020-uv}, Figure 9.}
\label{fig:pollen}
\end{figure}

Transformations of data are often useful prior to touring such as scaling, sphering, or a logarithmic transformation for skewed distributions. However, we may also want to transform the data after projecting onto lower dimensions, to correct for unwanted effects of the projections. One example is the piling effect that is observed when projecting a high-dimensional distribution \autocite{Diaconis1984-mv}: projected points are approximately Gaussian in most views, and increasingly concentrating near the center as dimensionality increases. The sage display \autocite{Laa2020-uv} proposes a solution to this issue, via a radial transformation of the projected data. This nonlinear transformation is defined in each point of the projection plane and is sensitive to the overall scale of the data as well as the original dimensionality \(p\). In addition, tuning parameters can be used to obtain a more (or less) aggressive redistribution of the points. This is illustrated in Figure \ref{fig:pollen} showing the application of the sage transformation to the classical pollen dataset \autocite{Coleman1986-so}. The projection without any rescaling looks similar to what is found when rescaling with default options (left), and we can use either of the two tuning parameters to better resolve the distribution near the center and reveal the hidden structure (middle and right).

\hypertarget{interaction}{%
\section{Ways of interacting}\label{interaction}}

\hypertarget{basic-interactions}{%
\subsection{Basic Interactions}\label{basic-interactions}}

\begin{figure*}
   \centering
         \includegraphics[width=0.9\textwidth]{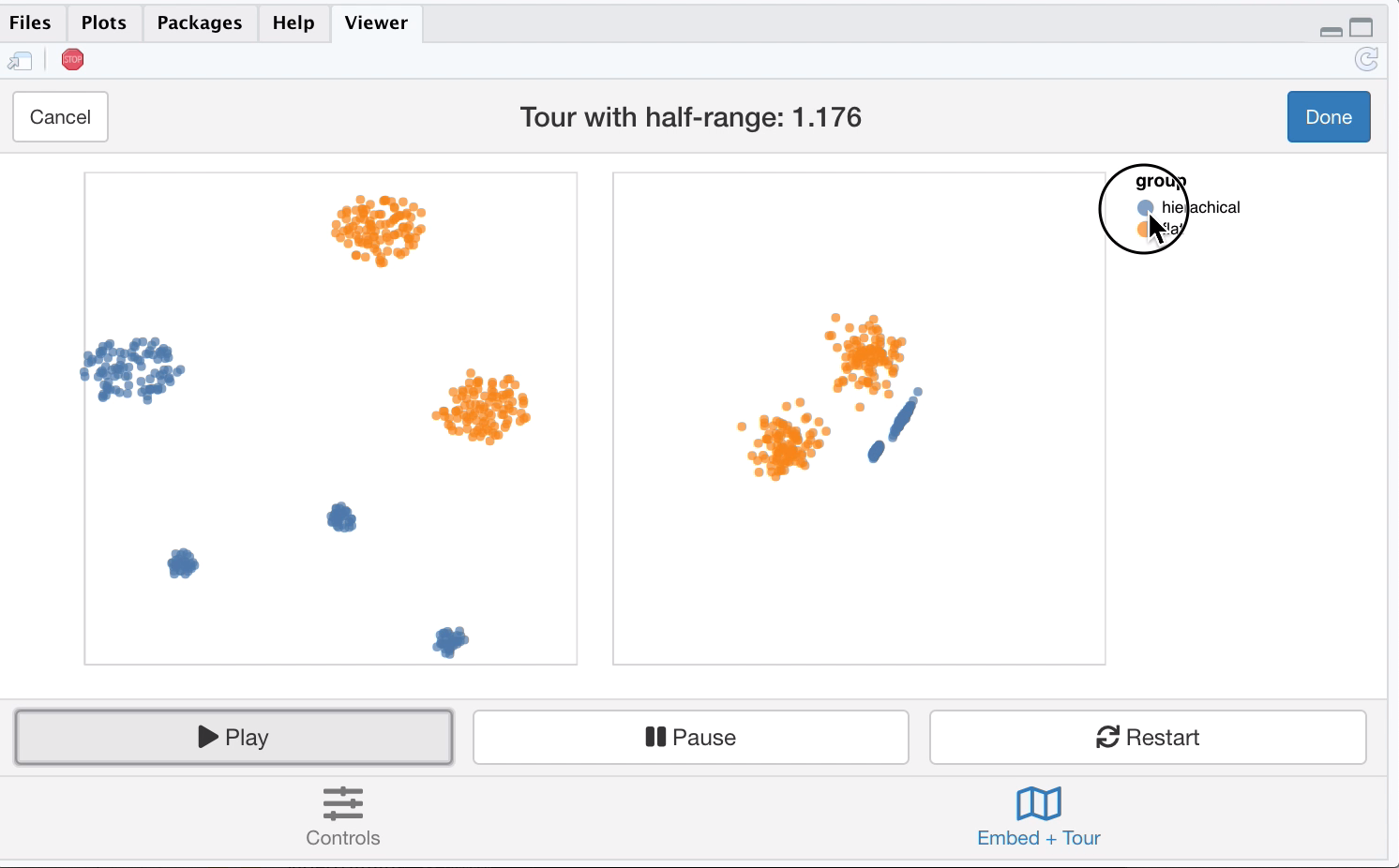}
\caption{The clustering example from Figure \ref{fig:tsne-tour} using the liminal interface. Here a grand tour is displayed on the right hand side, with buttons allowing users to play, pause and refresh the tour animation. Adapted from \cite{Lee2020-kg}, Figure 2.}
\label{fig:tour-controls}
\end{figure*}

Due to the dynamic nature of the tour, user interfaces can enhance the interpretability of the resulting visualization. Often one will want to pause on a particularly interesting projection and return it for use in a downstream analysis or static plot. Basic controls can be implemented so a tour display can be paused, refreshed and replayed over and over, as shown in Figure \ref{fig:tour-controls}.

\hypertarget{tours}{%
\subsection{Manual Tours}\label{tours}}

\begin{figure*}[!ht]
   \centering
         \includegraphics[width=0.32\textwidth]{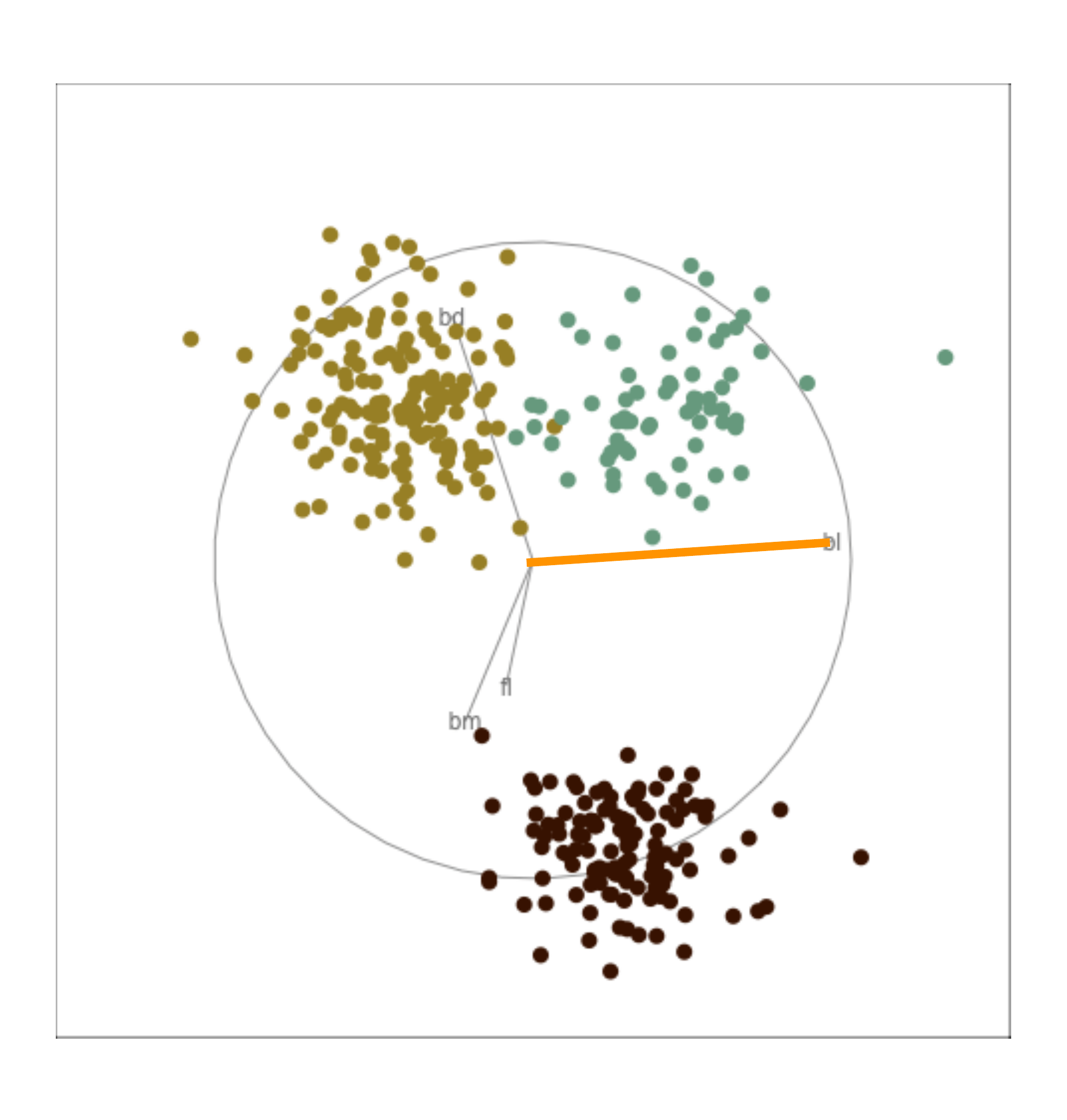}
         \includegraphics[width=0.32\textwidth]{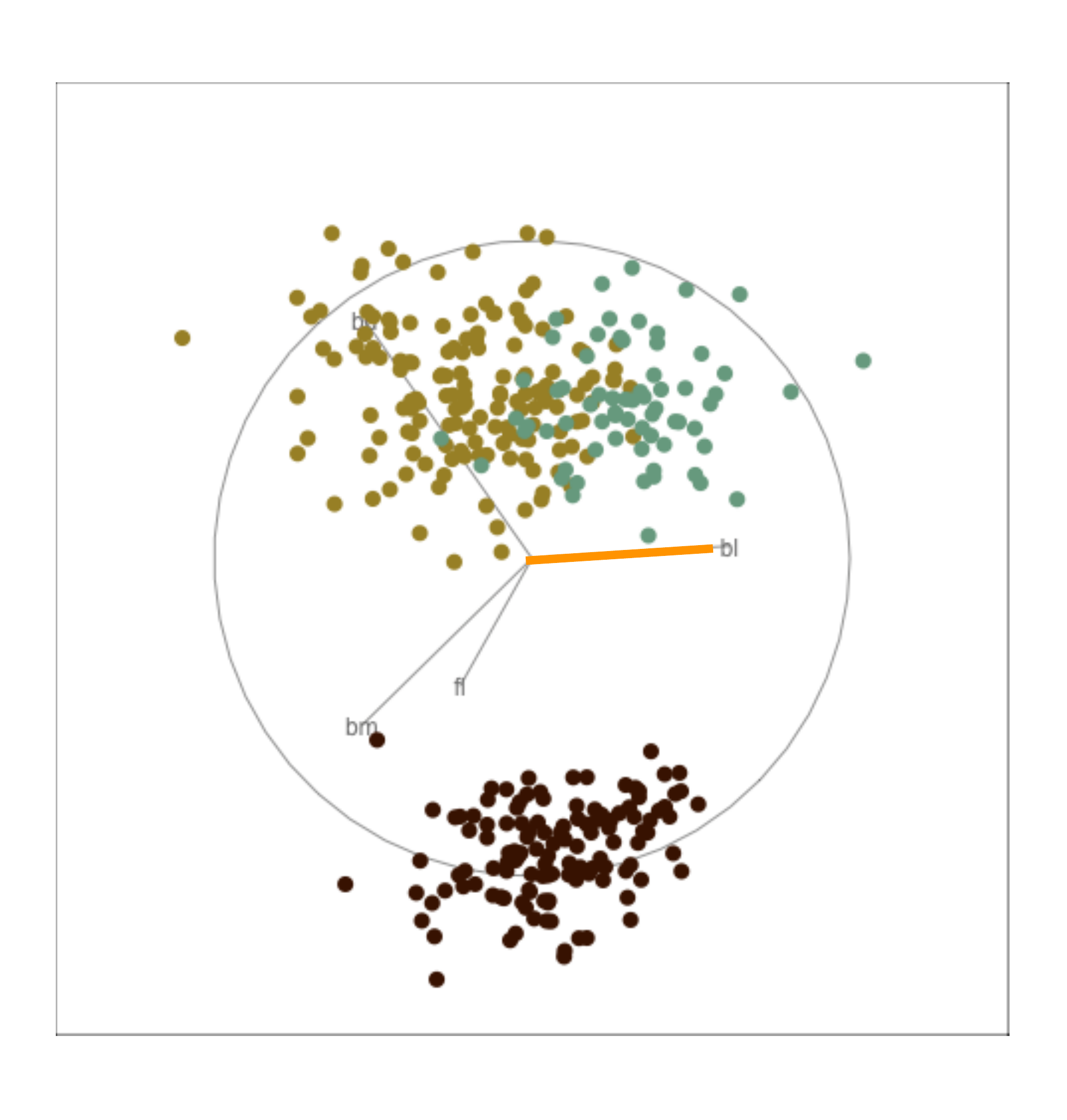}
         \includegraphics[width=0.32\textwidth]{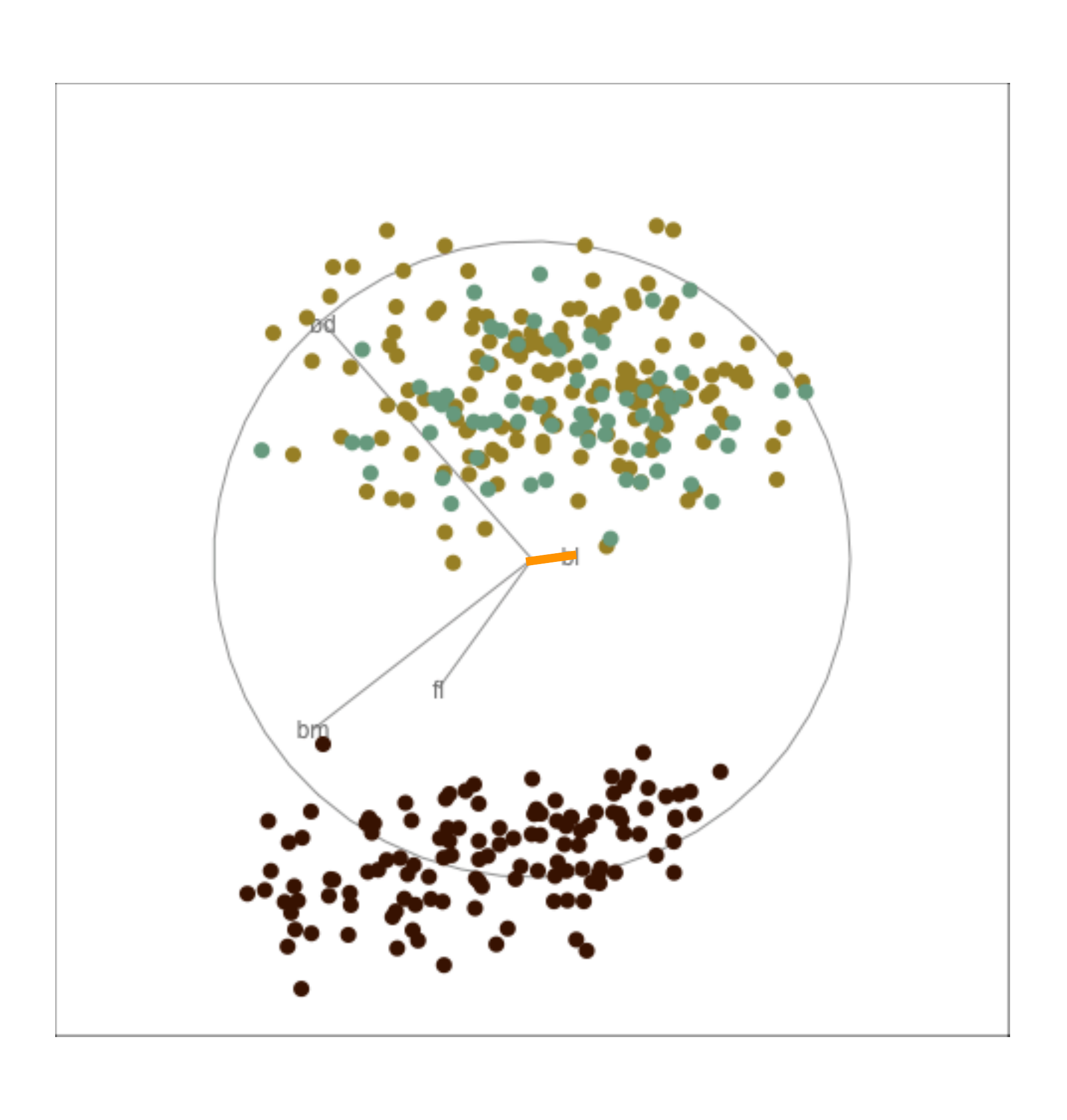}
\caption{Three projections from a manual tour where variable "bl" is being rotated out of the projection (orange line). When this variable is removed the two light green clusters merge, which informs us that "bl" is an important variable for distinguishing between these two groups.}
\label{fig:manual-tour}
\end{figure*}

Manual tours \autocite{cook_manual_1997,spyrison_spinifex_2020} offer a means to interactively control the contribution of a single variable on the projection plane. This is particularly useful for exploring a projection once a feature of interest has been identified. Manual tours can then be employed to test the structure of the feature, with respect to a selected variable. For instance, Figure \ref{fig:manual-tour}, starts from the orthonormal linear discriminant and explores the sensitivity of class separation as the contribution of a single variable is altered.

\hypertarget{sec:spin}{%
\subsection{Spin-and-brush}\label{sec:spin}}

\begin{figure}[htp]
\centerline{{\includegraphics[width=0.32\textwidth]{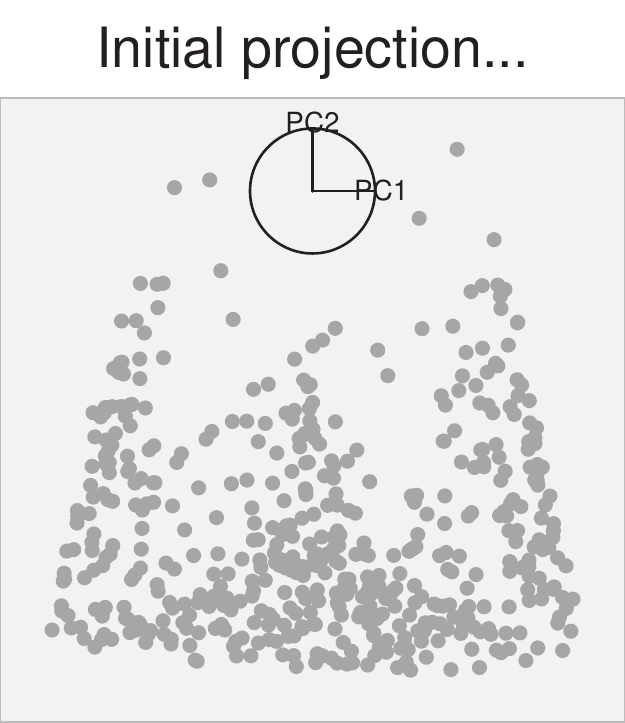}}
 {\includegraphics[width=0.32\textwidth]{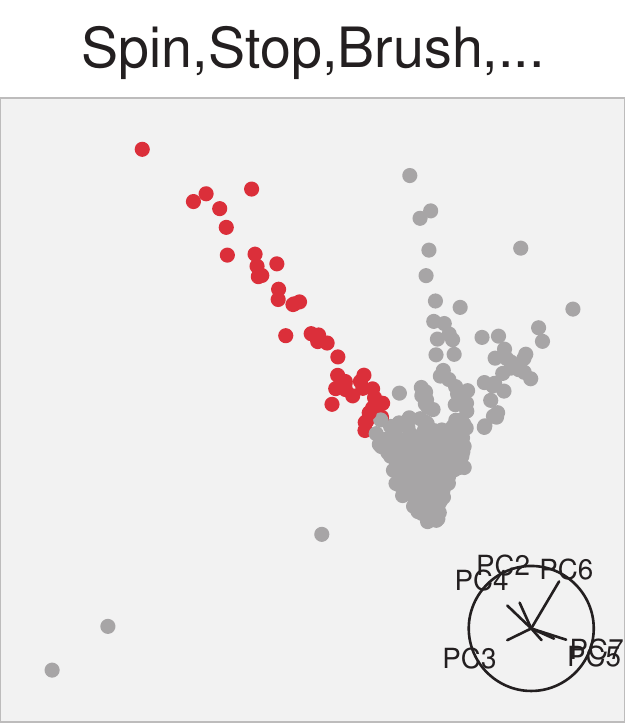}}
 {\includegraphics[width=0.32\textwidth]{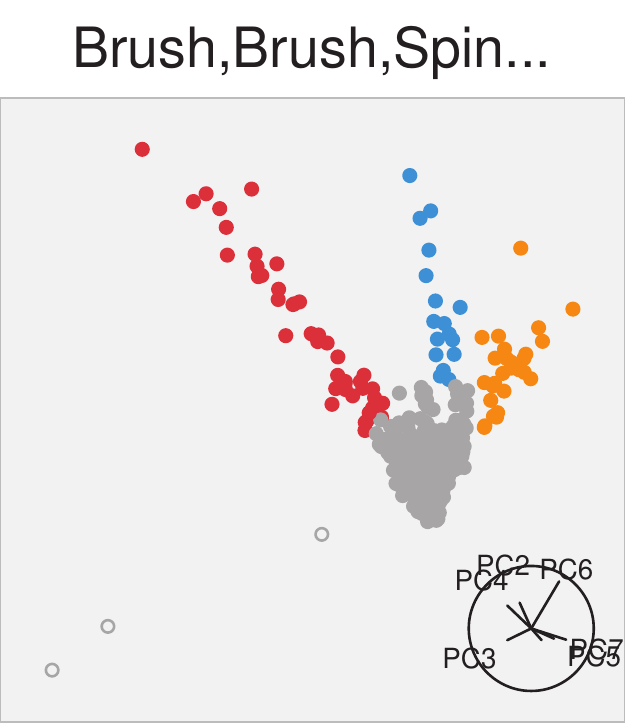}}}
\smallskip
\centerline{{\includegraphics[width=0.32\textwidth]{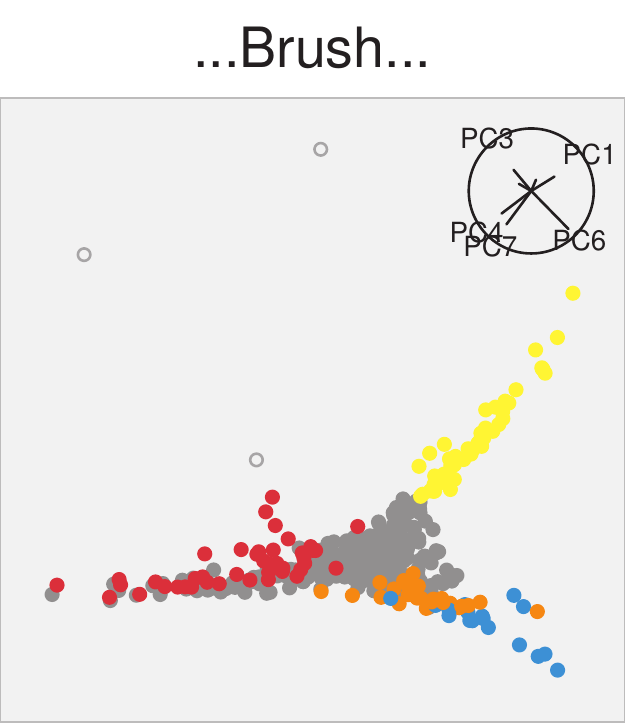}}
 {\includegraphics[width=0.32\textwidth]{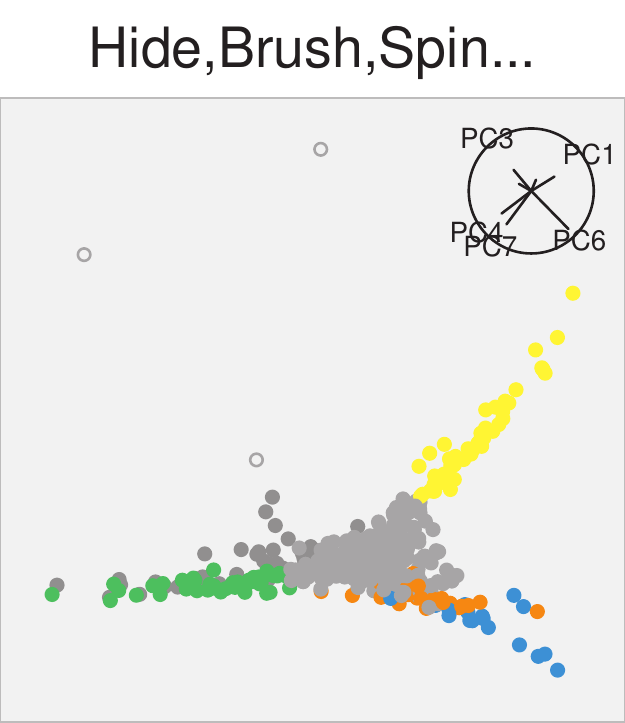}}
 {\includegraphics[width=0.32\textwidth]{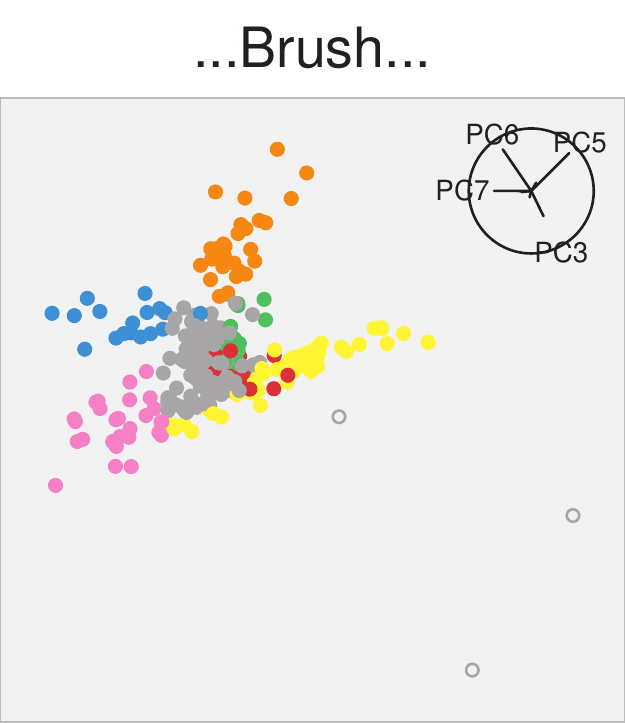}}}
\centerline{{\includegraphics[width=0.32\textwidth]{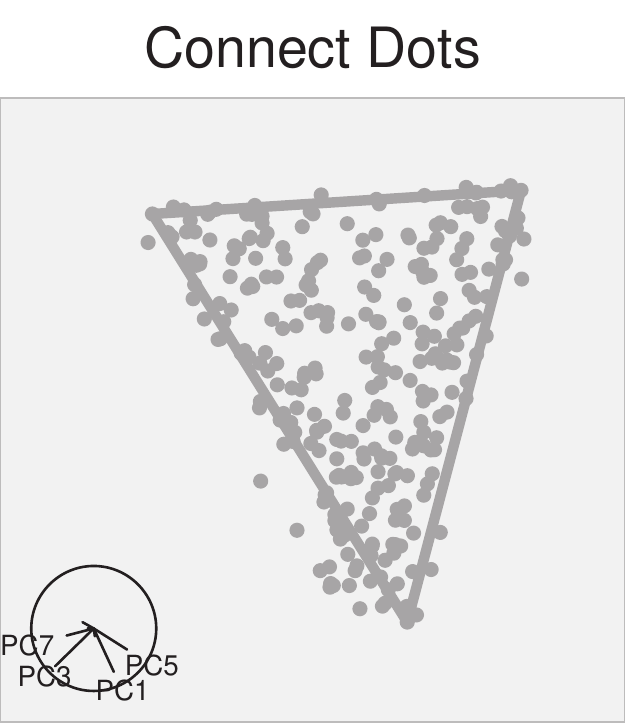}}
  {\includegraphics[width=0.32\textwidth]{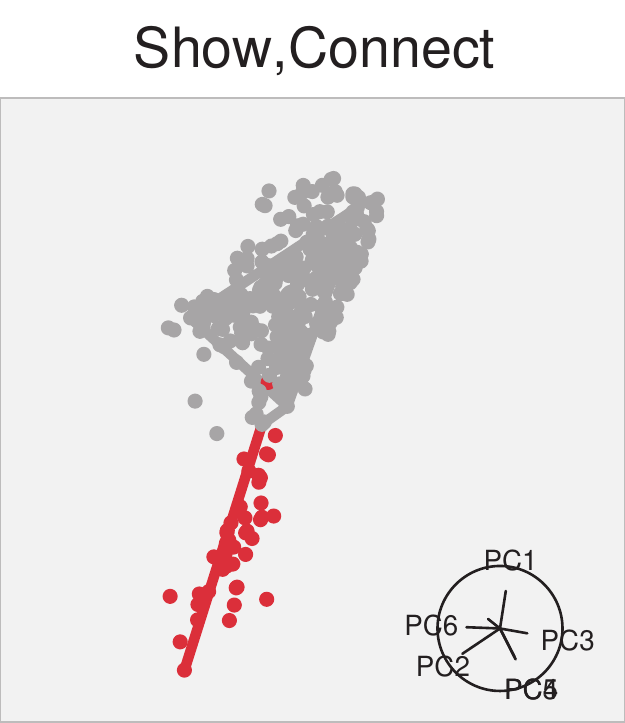}}
  {\includegraphics[width=0.32\textwidth]{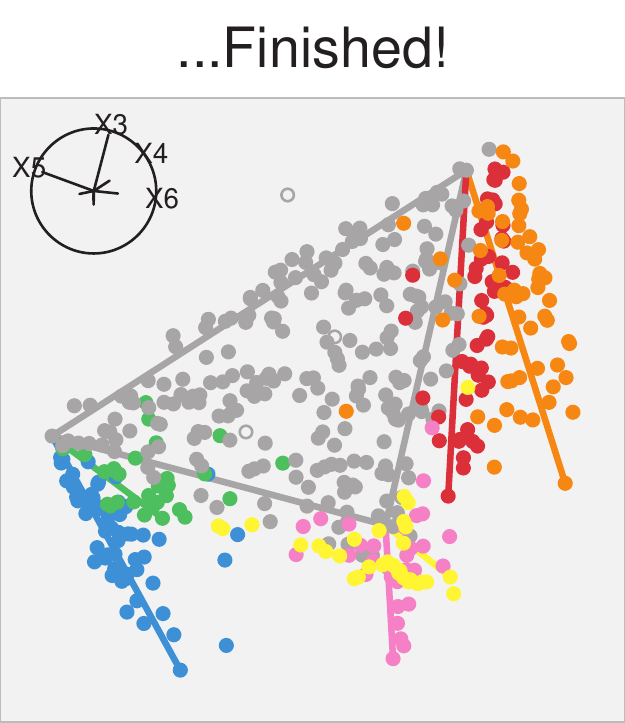}}}
\caption{Stepwise identification of clusters in the physics data. Each time a cluster is clearely separated we stop the tour and brush the points, followed by another spin of the data, until we can capture the full distribution. Adapted from \cite{Cook2007-be}, Figure 5.3.}
\label{fig:prim7-tour}
\end{figure}

Brushing can be used to aid statistical and geometrical interpretations of the data; brushing can be thought of as conditioning variables on certain regions of the data or used to section lower dimensional views. This aids tasks such as the identification of outlying points or visual cluster analysis when combined with the tour. Here, a ``spin-and-brush'' approach works well, since different views will reveal different features and a persistent brush helps us to connect this to the previously observed information. As an example we briefly summarize an analysis from \textcite{Cook2007-be} that explores clustering of the physics data previously used to in the context of projection pursuit \autocite{Friedman1974-ck,Cook1995-ae}. This data is well described by its geometric structure: a two-dimensional triangle with two one-dimensional strands linearly extending in different directions from each vertex. These strands can sequentially be identified as clusters using the spin-and-brush approach, as shown in Figure \ref{fig:prim7-tour}. We stop the tour each time a cluster is clearly separate from the main distribution in the current projection and brush the points in a new color. In the end the full structure becomes apparent and can be visualized by replaying the tour with all clusters highlighted in different colors.

\hypertarget{software}{%
\section{Software}\label{software}}

There has been a long history of software including tour methods, or something similar. These have laid the foundation for the current tools. The American Statistical Association Statistical Graphics Video Library web site \autocite{ASASGVL} hosts videos documenting the history. It is fabulous watching the videos titled ``Multidimensional Scaling'' (Kruskal, 1962), ``Real-Time Rotation'' (Chang, 1970) and ``Prim-9'' (Tukey, 1973) show some preliminary methods leading to the development of the grand tour. The video titled ``Use of the Grand Tour in Remote Sensing'' (McDonald and Willis, 1987) is the first to show a tour, and the video titled ``Dataviewer: A Program for Looking at Data in Several Dimensions'' (Buja and Tukey, 1987) demonstrates the tour as part of a larger data analysis system. The videos titled ``XGobi: Dynamic Graphics for Data Analysis'' (Swayne et al, 1991) and ``Grand Tour and Projection Pursuit'' (Cook et al, 1993) show the tour tools in the XGobi software system. Some work on tours, for example \textcite{Wegman92} and \textcite{Tierney1991} is not documented in the video collection.

Recently, some other review papers have summarized different aspects of tour methods. \textcite{Moustafa2009} explains the relationship between a grand tour and the classical multivariate plot, the Andrews curve. \textcite{Moustafa2010} describes a computational shortcut for a tour. \textcite{Martinez2013} explains the image grand tour.

The most accessible current software is the R package called tourr \autocite{Wickham2011-uz}. This software evolved from the Dataviewer \autocite{Buja1986-la}, XGobi \autocite{Swayne1998}, GGobi \autocite{Swayne2003}, Orca \autocite{Sutherland2000} and cranvas \autocite{Xie2014} ancestry. The tourr package has a wide range of display types, for different projection dimensions, and a selection of target generating methods including grand, guided, little, local, section, sage and frozen.

The R package spinifex \autocite{spyrison_spinifex_2020} has 3 primary features: it produces manual tours (predefined path or interactive manipulation, identifies orthonormal global feature bases with the use of the Rdimtools \autocite{you_rdimtools_2020} package, renders (manual or other) tours as animations exportable to static .gif (gganimate \autocite{pedersen_gganimate_2020} package) or interactive .html widgets (plotly \autocite{sievert_interactive_2020} package), lastly it offers interactive shiny application that offers an graphical user interface to quickly sample tour features.

The liminal R package uses the tour to explore the quality of non-linear dimensionality reduction algorithms \autocite{Lee2020-kg}. The interface consists of two side by side views consisting of a scatter plot displaying a reduced form of the data, and an interactive tour. Controls such as play/stop/restart are implemented allowing a user to pause on interesting projections and return them to their R session for further analysis. Linked brushing is implemented on both views - if a users brush on the scatter plot view they can see if and how an algorithm like t-SNE has distorted distances in higher dimensional space, while if they brush on the tour view, the tour is paused and structure like cluster separation can be ascertained.

A Python implementation of the tour written by \textcite{Nguyen2020} is also available, and would benefit from extension by other developers.

\hypertarget{applications}{%
\section{Applications}\label{applications}}

\hypertarget{model-visualization}{%
\subsection{Model Visualization}\label{model-visualization}}

\begin{figure*}
  \centering
    \includegraphics[width=0.32\textwidth]{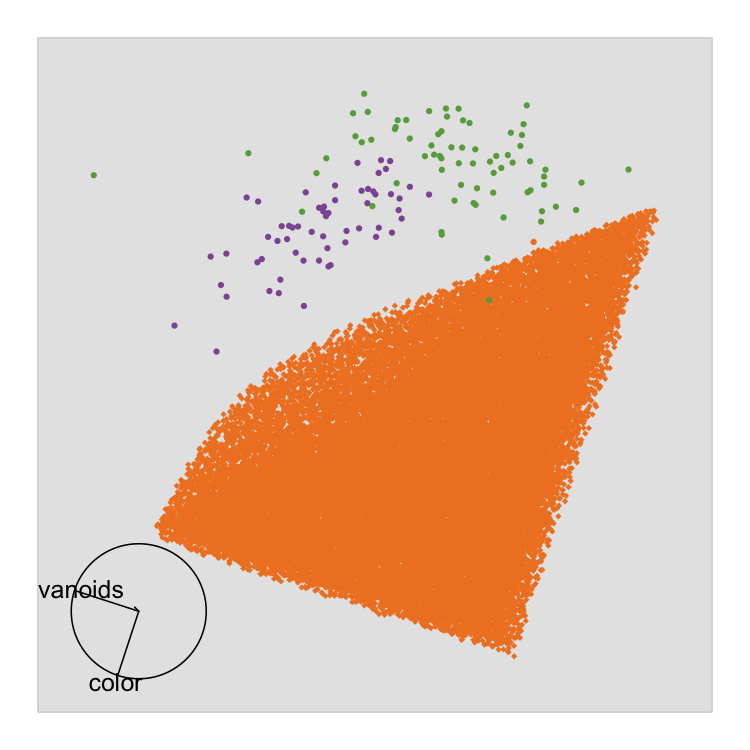}
    \includegraphics[width=0.32\textwidth]{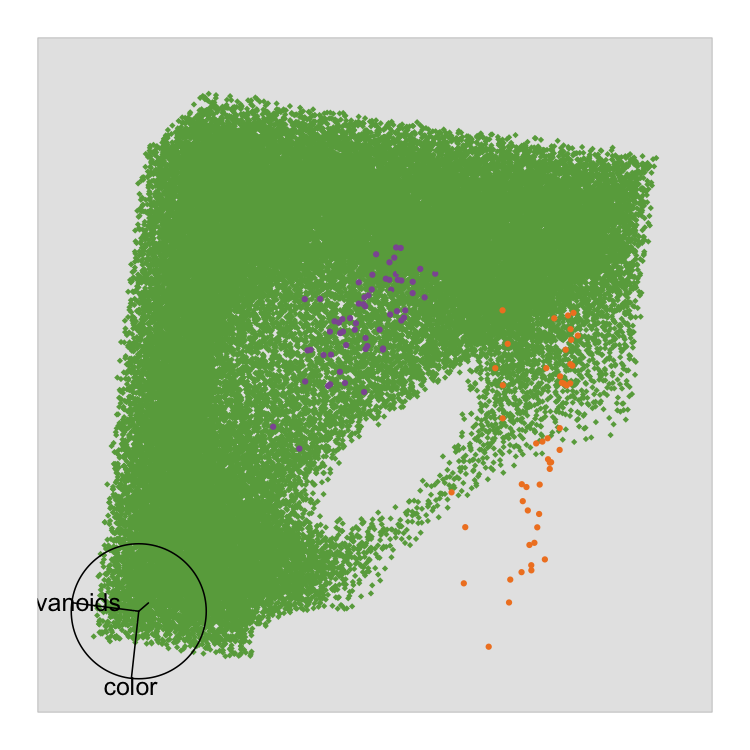}
    \includegraphics[width=0.32\textwidth]{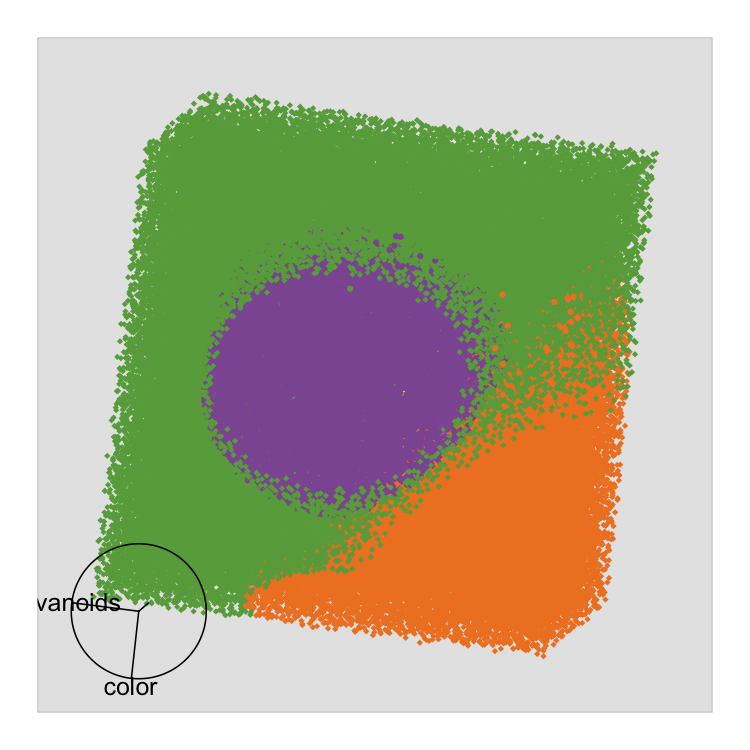}
  \caption{A grand tour run on the prediction boundaries of a radial support vector machine. (Left) The prediction region for points classified as orange produces a slice of a cube. (Middle) The predicition region for points classified as green produces another slice of the cube. (Right) The complete prediction region shows the points classified as purple are embedded within the green slice. Adapted from \cite{Wickham2015-do}, Figure 6.}
  \label{fig:svm-tour}
\end{figure*}

\textcite{Wickham2015-do} espouses the tour as a key component of their approach for visualizing models in data space. They propose using the tour to explore the high dimensional surfaces produced by model fits instead of simple summary statistics. In Figure \ref{fig:svm-tour} samples have been taken along the prediction region (i.e.~the value the model predicts over a grid of observations) of a 3-\(d\) support vector machine used to classify three different classes (green, orange purple), by touring over the prediction region we can see how the classifier splits the decision boundaries.

\begin{figure*}
   \centering
         \includegraphics[width=0.9\textwidth]{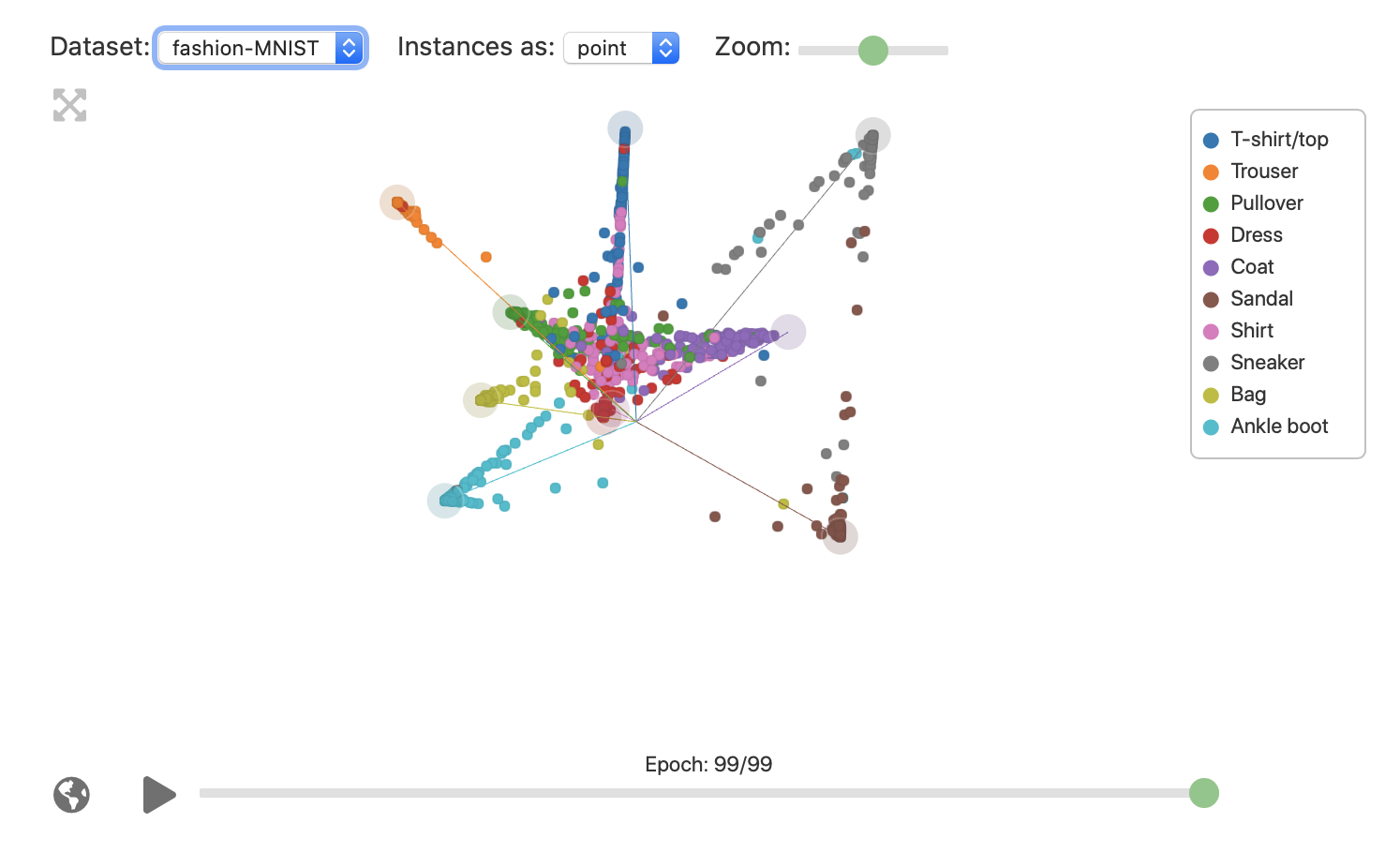}
\caption{A grand tour run on the output of a deep neural network model for classification. The axes correspond to the probabilities that the model of an image belonging to a certain class. Adapted from \cite{Li2020-kg}.}
\label{fig:tour-nn}
\end{figure*}

The grand tour along with direct interaction techniques has been proposed as a technique for understanding the training of deep neural network models \autocite{Li2020-kg}. They proposed a novel method for aligning the output of different layers by interpolating the output of different layers of a deep neural network via the the grand tour. They also apply the grand tour to explore the changes in single layers of the network for each training epoch. For example, by the running tour over the softmax layer outputs of a classification model, an analyst may understand where in the training confusion between classes occur and gain a view of model performance (Figure \ref{fig:tour-nn}). Similarly, the difference between training and test sets can be assessed through side by side linked tours of layer outputs. \textcite{li_toward_2020} extends this idea by first reducing the dimensionality of output layers with UMAP \autocite{mcinnes2020umap} and then using the grand tour with manual controls over 15 dimensional embeddings to understand layer topology.

\hypertarget{physics}{%
\subsection{Physics}\label{physics}}

\begin{figure*}
  \centering
           \includegraphics[width=0.45\textwidth]{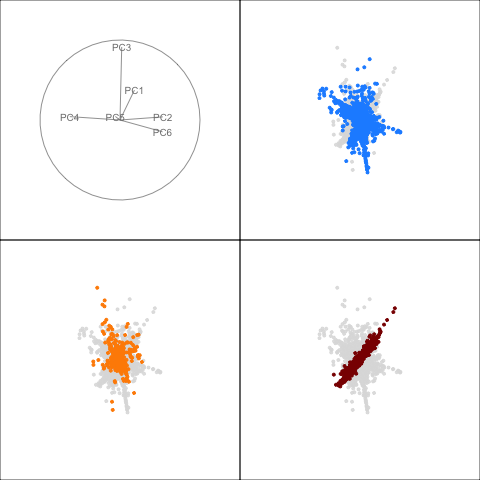}
         \includegraphics[width=0.45\textwidth]{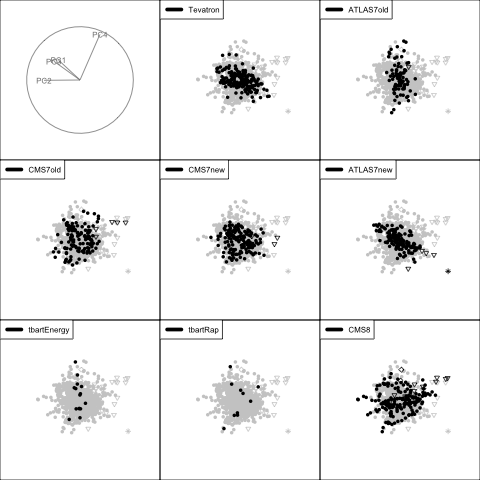}
\caption{Selected frames from using a grand tour of the physics data. Left: the three different types of measurements (shown in different colors) are aligned along different directions in parameter space. Right: an outlying point is marked with an asterisk symbol in the ATLAS7new facet. Adapted from \cite{Cook2018-jm}, Figures 6 and 7.}
\label{fig:physics}
\end{figure*}

Physics models often consider \(\mathcal{O}(10)\) free parameters and are compared to a much larger number of experimental observables (\(\mathcal{O}(100)\)). The comparison typically relies on numeric computation of the model predictions for all observables, potentially obscuring the nature of their dependence on the model parameters. Here, multivariate visualization methods can provide new insights. This was demonstrated in \textcite{Cook2018-jm}. For example, we explored grouping of experiments based on how they constrain the parameter space and we also identified an interesting multivariate outlier that pointed to potential issues with the data point. For illustration we show static views from the tour showing the orthogonal structure of the three groups (indicated by color), and a display which highlights the ``outlyingness'' of the data point marked with an asterisk symbol in Figure \ref{fig:physics}.

\hypertarget{bioinformatics}{%
\subsection{Bioinformatics}\label{bioinformatics}}

In single cell RNA sequencing, scientists are interested in identifying novel cell types and understanding the relationships between cells or their developmental trajectory. To achieve this they perform cluster analysis on a counts matrix or principal components and embed the results via t-SNE and label points according the cluster label. One of the main advantages of t-SNE is the avoidance of over-plotting so clusters can be clearly identified on a scatter plot, however, this can come at the cost of interpretability as global distances are distorted. In \textcite{Laa2020-uv}, we used radial transformations of the tour projections as an alternative to t-SNE that better preserves global structure while still retaining cluster topology.

\begin{figure*}
  \centering
           \includegraphics[width=0.24\textwidth]{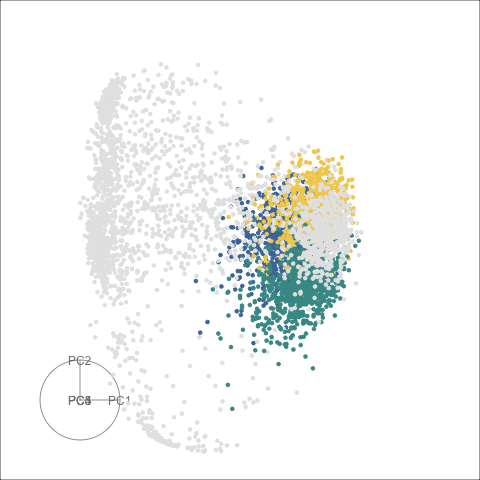}
         \includegraphics[width=0.24\textwidth]{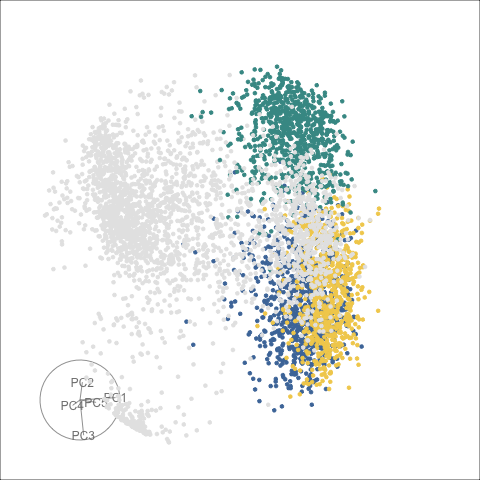}
         \includegraphics[width=0.24\textwidth]{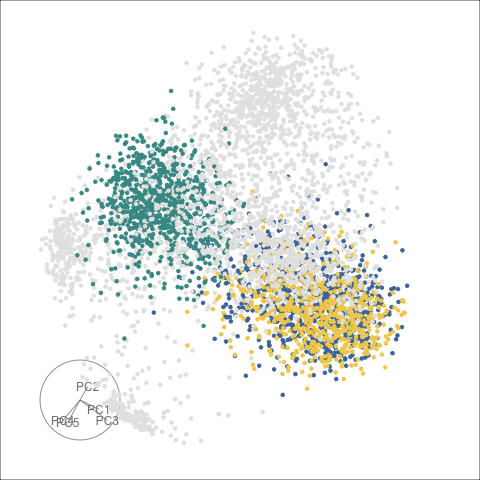}
         \includegraphics[width=0.24\textwidth]{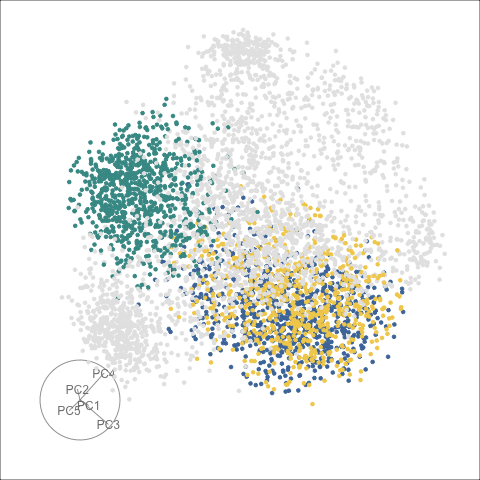}

\caption{Selected frames from using a grand tour of mouse retina single cell RNA-seq data with the radial transformation. The sage display was used to identify and verify cluster separation between the three highlighted clusters estimated from a clustering algorithm. Adapted from \cite{Laa2020-uv}, Figure 6.}
\label{fig:mouse-cluster}
\end{figure*}

\hypertarget{geometry-of-data}{%
\subsection{Geometry of data}\label{geometry-of-data}}

\begin{figure*}
  \centering
  \includegraphics[width=0.24\textwidth]{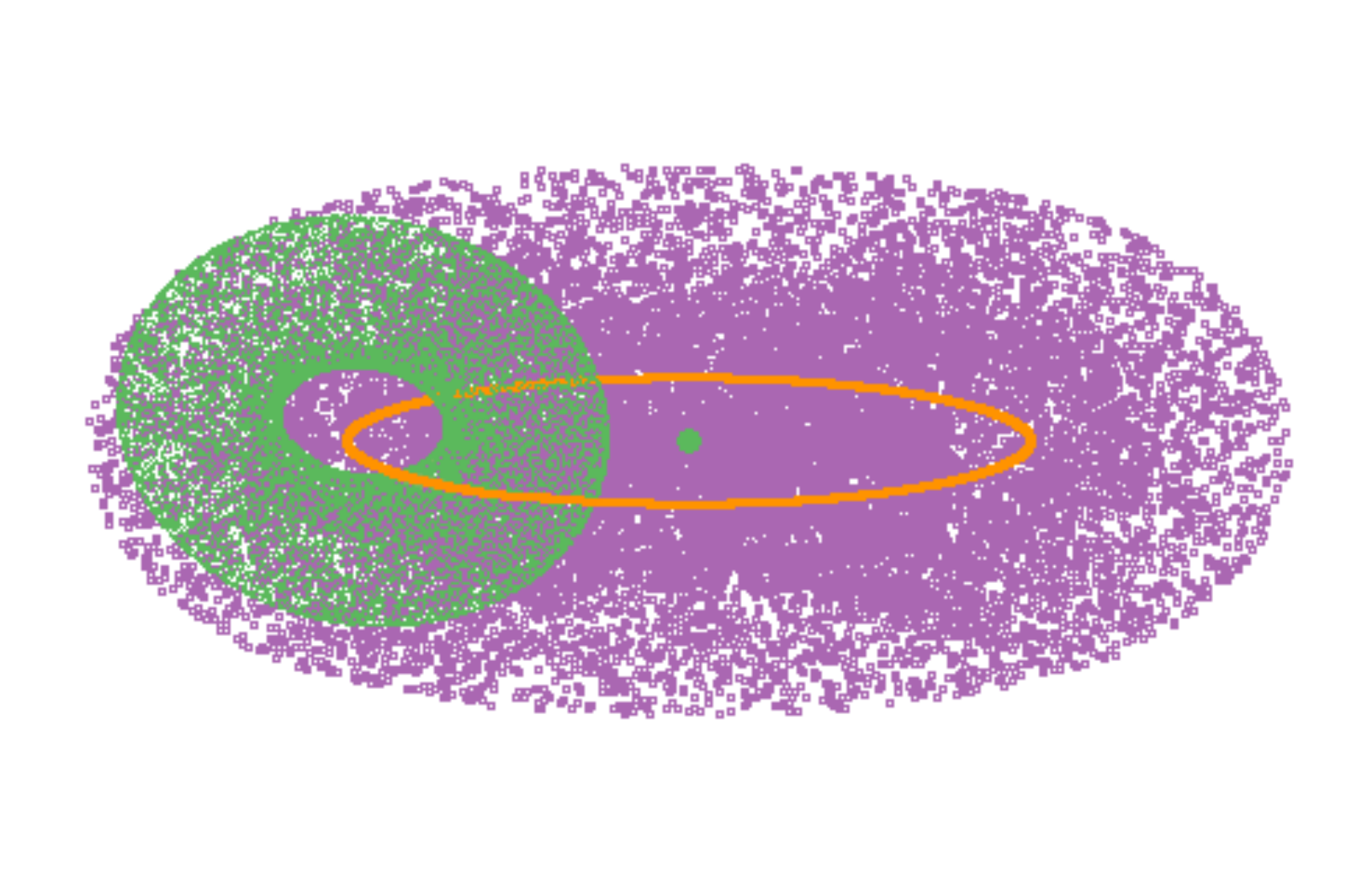}\includegraphics[width=0.24\textwidth]{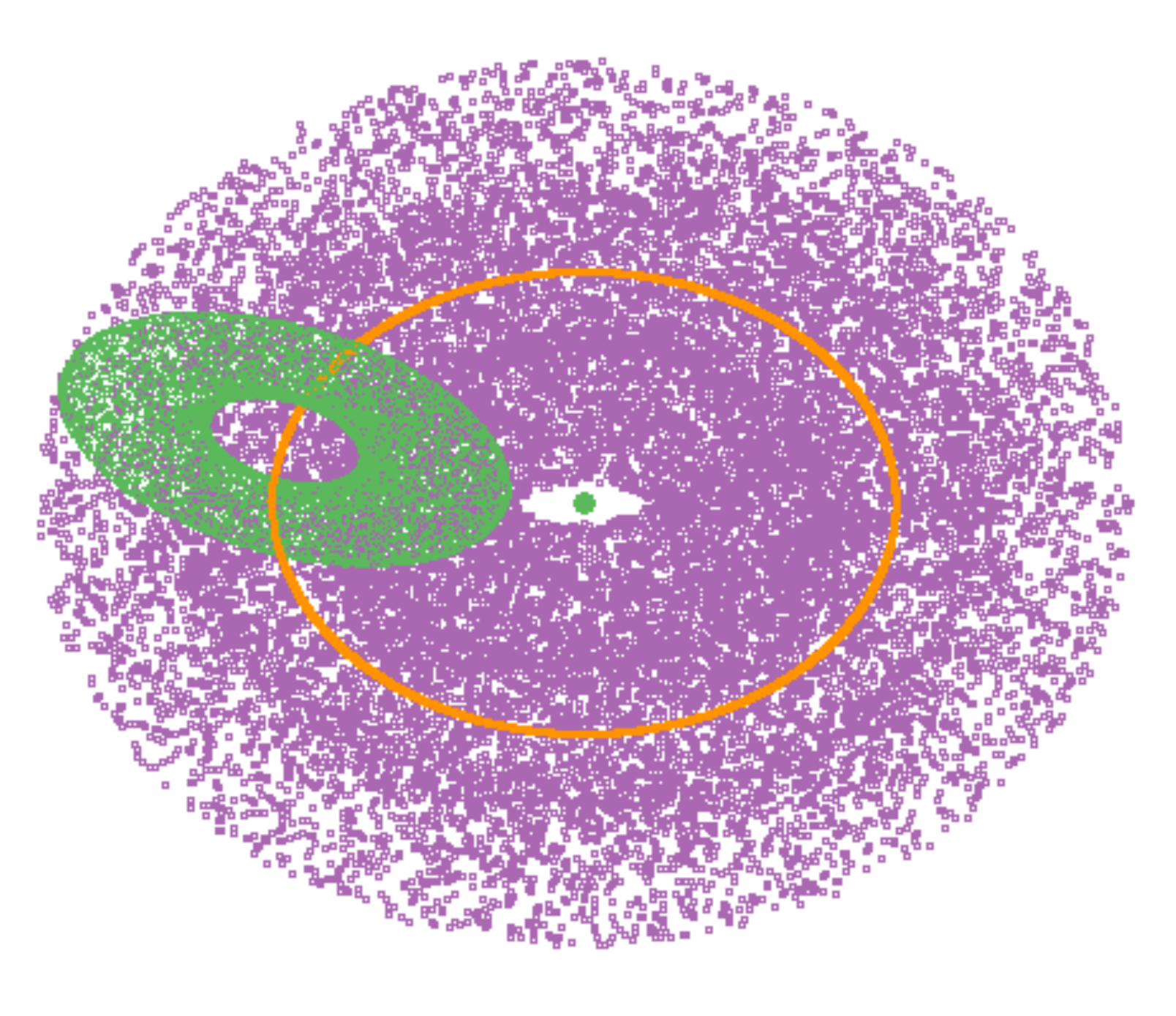}
  \includegraphics[width=0.24\textwidth]{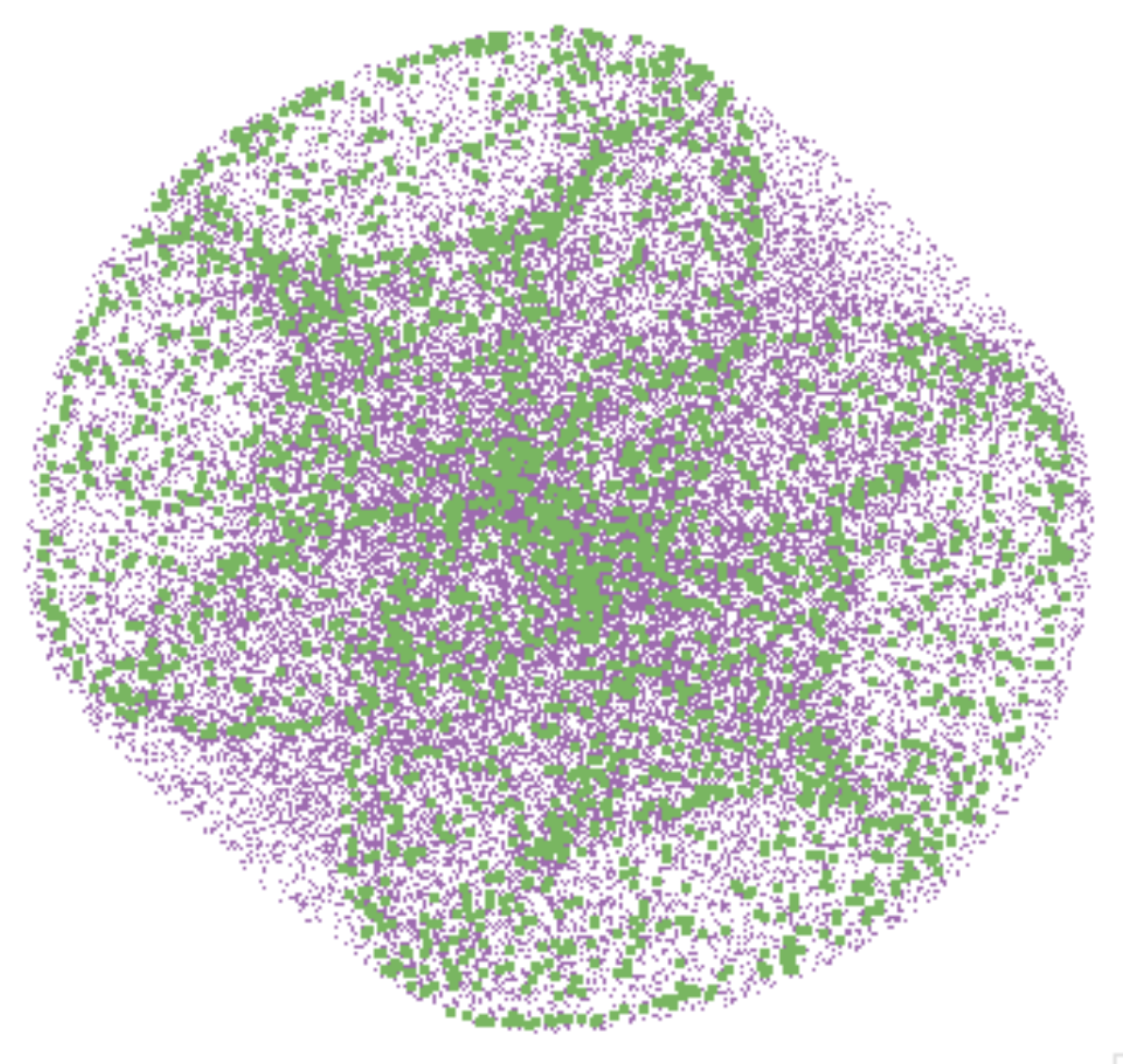}\includegraphics[width=0.24\textwidth]{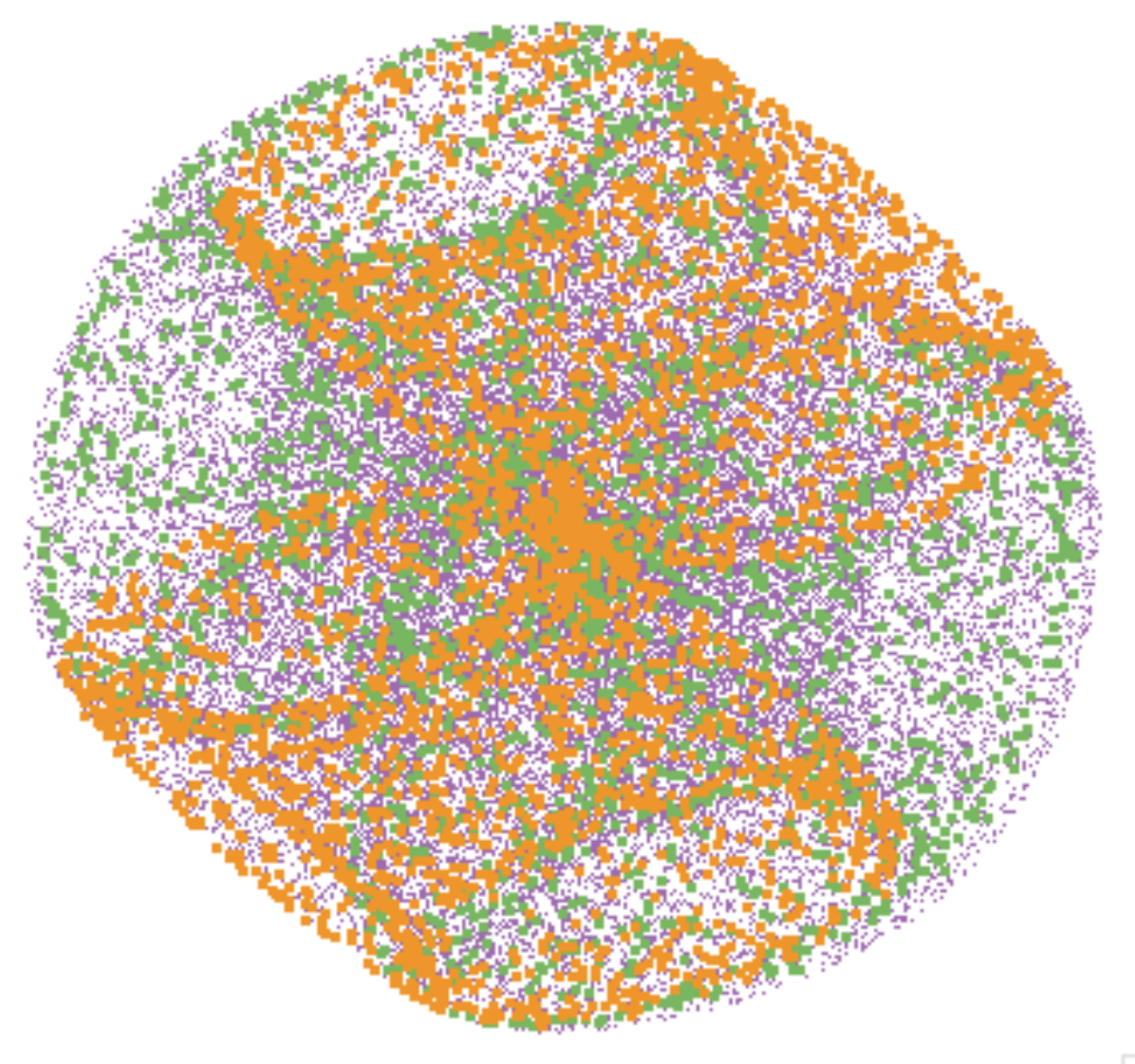}
\caption{Several views of torii: (two at left) 4D ring torus, and (two at right) 6D flat torus. In each case, a subset has been highlighted which illustrate the low-dimensional donut shapes. Adapted from \cite{RJ-2016-044}, Figures 10 and 13.}
\label{fig:geozoo}
\end{figure*}

High-dimensional data is on many analysts' minds in recent years, and a way to build understanding of high-dimensional spaces is to examine common high-dimensional shapes using a tour. Many different shapes can be simulated using the R package geozoo \autocite{RJ-2016-044}. Figure \ref{fig:geozoo} shows points on the surface of two different types of high-dimensional torii.

\hypertarget{discussion}{%
\section{Discussion}\label{discussion}}

This review highlights the history of tour methods, and modern extensions for the visualization of high dimensional numerical data. The rich ecosystem of tours is being actively developed and applied to a broad range of problems including machine learning. When tours are be combined with some interactivity it can assist with analyses like clustering (the spinifex and liminal packages). Modified displays of projected data can be used to expose local structures (the sage algorithm, slicing and section pursuit). New algorithm data collection inside the tour code is being used to better diagnose optimization procedures for projection pursuit \autocite{zhang2021visual}.

There are many possible future directions for tour research. One big challenge for software engineering is to seamlessly embed the tour with interactivity. Currently, there is no implementation that analysts can use to perform tasks that were easy in software available in the 80s and 90s. More contributions in sectioning as an alternative to projection, could be broadly useful. An example is conditional model visualization, as implemented in the condviz package \autocite{JSSv081i05}, which assists with visual exploration of multivariate fitted models. It could be interesting to define tour paths for different types of nonlinear subspaces, or non-Euclidean space.

\hypertarget{acknowledgements}{%
\section*{Acknowledgements}\label{acknowledgements}}
\addcontentsline{toc}{section}{Acknowledgements}

The authors gratefully acknowledge the support of the Australian Research Council. The paper was written in \texttt{rmarkdown} \autocite{Xie2018-bg} using \texttt{knitr} \autocite{Xie2017-uh}. The source material for this paper is available at \url{https://github.com/dicook/wiley-isghdd}.

\printbibliography[title=References]

\end{document}